\documentclass[aps,prb,preprint,preprintnumbers,amsmath,amssymb,superscriptaddress]{revtex4}
\usepackage{graphicx}
\usepackage{bm}
\usepackage[final]{pdfpages}

\begin{document}
\title{Majorana fermions in a superconducting M\"{o}bius strip}
\date{\today}

\author{Yuan Pang}\thanks{These authors contribute equally to this work.} \affiliation{Beijing National Laboratory for Condensed Matter Physics, Institute of Physics, Chinese Academy of Sciences, Beijing 100190, People's Republic of China}
\author{Jie Shen}\thanks{These authors contribute equally to this work.}
\affiliation{Beijing National Laboratory for Condensed Matter Physics, Institute of Physics, Chinese Academy of Sciences, Beijing 100190, People's Republic of China}
\author{Fanming Qu} \thanks{Present address: QuTech and Kavli Institute of Nanoscience, Delft University of Technology, 2600 GA Delft, The Netherlands.}
\affiliation{Beijing National Laboratory for Condensed Matter Physics, Institute of Physics, Chinese Academy of Sciences, Beijing 100190, People's Republic of China}
\author{Zhaozheng Lyu} \affiliation{Beijing National Laboratory for Condensed Matter Physics, Institute of Physics, Chinese Academy of Sciences, Beijing 100190, People's Republic of China}
\author{Junhua Wang} \affiliation{Beijing National Laboratory for Condensed Matter Physics, Institute of Physics, Chinese Academy of Sciences, Beijing 100190, People's Republic of China}
\author{Junya Feng} \affiliation{Beijing National Laboratory for Condensed Matter Physics, Institute of Physics, Chinese Academy of Sciences, Beijing 100190, People's Republic of China}
\author{Jie Fan} \affiliation{Beijing National Laboratory for Condensed Matter Physics, Institute of Physics, Chinese Academy of Sciences, Beijing 100190, People's Republic of China}
\author{Guangtong Liu} \affiliation{Beijing National Laboratory for Condensed Matter Physics, Institute of Physics, Chinese Academy of Sciences, Beijing 100190, People's Republic of China}
\author{Zhongqing Ji} \affiliation{Beijing National Laboratory for Condensed Matter Physics, Institute of Physics, Chinese Academy of Sciences, Beijing 100190, People's Republic of China}
\author{Xiunian Jing} \affiliation{Beijing National Laboratory for Condensed Matter Physics, Institute of Physics, Chinese Academy of Sciences, Beijing 100190, People's Republic of China} \affiliation{Collaborative Innovation Center of Quantum Matter, Beijing 100871, People's Republic of China}
\author{Changli Yang} \affiliation{Beijing National Laboratory for Condensed Matter Physics, Institute of Physics, Chinese Academy of Sciences, Beijing 100190, People's Republic of China} \affiliation{Collaborative Innovation Center of Quantum Matter, Beijing 100871, People's Republic of China}
\author{Qingfeng Sun} \affiliation{Collaborative Innovation Center of Quantum Matter, Beijing 100871, People's Republic of China} \affiliation{International Center for Quantum Materials, Peking University, Beijing 100871, People's Republic of China}
\author{X. C. Xie} \affiliation{Collaborative Innovation Center of Quantum Matter, Beijing 100871, People's Republic of China} \affiliation{International Center for Quantum Materials, Peking University, Beijing 100871, People's Republic of China}
\author{Liang Fu} \affiliation{Department of Physics, Massachusetts Institute of Technology, Cambridge, Massachusetts 02139, USA}
\author{Li Lu} \email[Corresponding author. Email: ]{lilu@iphy.ac.cn} \affiliation{Beijing National Laboratory for Condensed Matter Physics, Institute of Physics, Chinese Academy of Sciences, Beijing 100190, People's Republic of China} \affiliation{Collaborative Innovation Center of Quantum Matter, Beijing 100871, People's Republic of China}

\begin{abstract}
Recently, much attention has been paid to search for Majorana fermions in solid-state systems. Among various proposals there is one based on radio-frequency superconducting quantum interference devices (rf-SQUIDs), in which the appearance of 4$\pi$-period energy-phase relations is regarded as smoking-gun evidence of Majorana fermion states. Here we report the observation of truncated 4$\pi$-period (i.e., 2$\pi$-period but fully skewed) oscillatory patterns of contact resistance on rf-SQUIDs constructed on the surface of three-dimensional topological insulator Bi$_2$Te$_3$. The results reveal the existence of $1/2$ fractional modes of Cooper pairs and the occurrence of parity switchings, both of which are necessary signatures accompanied with the formation of Majorana fermion states.
\end{abstract}

\begin{figure}[b]
\vspace{1 cm}
\includegraphics[width=0.8 \linewidth]{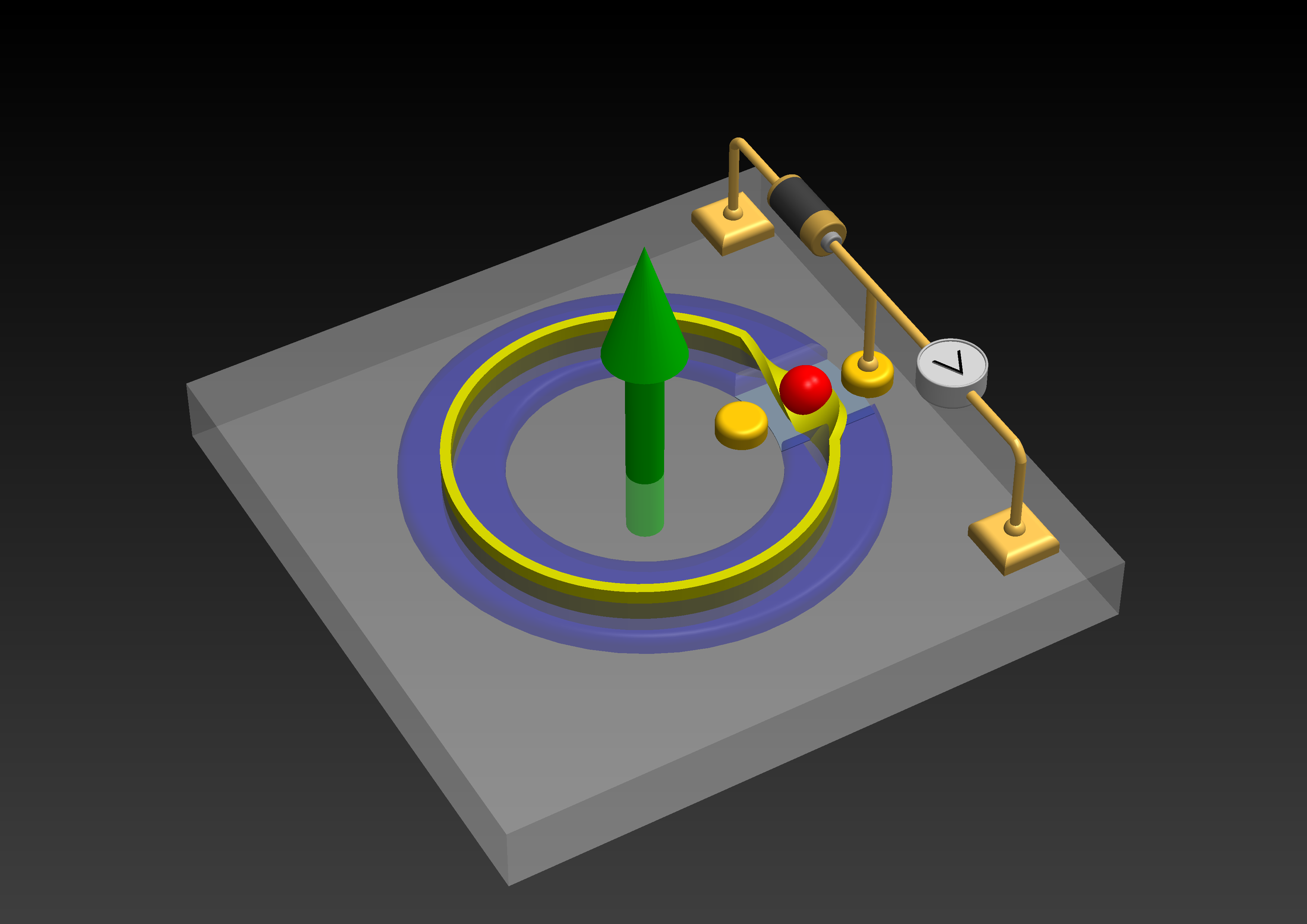}

\label{cover}{}
\end{figure}

\maketitle

Majorana Fermions are a mysterious type of particles predicted eight decades ago \cite{Majorana} but are still nowhere to be found. Recently there is a hope to find Majorana quasiparticles in condensed matter systems \cite{Majorana_return_Wilczek,Majorana_return_Service,Majorana_return_Franz}. The pairing of helical electrons in topological insulators (TIs) and related materials are expected to form topological superconductivity resembling that of a spinless $p$-wave superconductor, and hosting Majorana fermion states \cite{Fu_liang_2008} with which topological quantum computers could potentially be built. One of the signatures of Majorana fermion states, the appearance of a zero-bias conductance peak at the superconductor-normal metal (S-N) interface, has already been observed experimentally \cite{Kowenhoven}. Also highly-sought is the appearance of 4$\pi$-period energy-phase relations (EPRs) in S-TI-S type of Josephson junctions \cite{Kitaev2001,Fu_Kane_2009ring,Lutchyn_Sau_Das_Sarma_2010ring,Beenakker_2013ring,Benjamin_Zhang_Kane_2014ring}. To search for the 4$\pi$-period EPRs, a number of phase-sensitive experiments have been conducted on various Josephson devices constructed on 2D and 3D TIs \cite{Brinkman,QuFM,Goldhaber-Gordon,Rokhinson,HarlingenTI,Moler_PRL,Tarucha_Molenkamp,Yacoby_NW,Yacoby_QPP,Kouwenhoven_Qu}, but clear evidence is still awaited.

In 2008, Fu and Kane \cite{Fu_liang_2008} proposed that Majorana fermion states should exist in Josephson junctions constructed on the surface of a 3D TI, as long as the phase difference across the junction is $\pi$. To maintain the $\pi$ phase difference, an ingenious way is to make a rf-SQUID by connecting the Josephson junction with a superconducting ring (Fig. 1) and threading half flux quantum into the ring. Such an experiment has been proposed in several different versions by the theorists \cite{Fu_Kane_2009ring,Lutchyn_Sau_Das_Sarma_2010ring,Beenakker_2013ring,Benjamin_Zhang_Kane_2014ring}. For rf-SQUIDs constructed on the surface of 3D TIs, in particular, Wieder, Zhang and Kane predicted that the existence of Majorana fermion states in the junction would accompany with a 4$\pi$-period signature in the channel conductance of the junction, resulting in quantum phase transitions and conductance jumpings at every odd multiples of half flux quantum \cite{Benjamin_Zhang_Kane_2014ring}. In such case, we anticipate that the local conductance spectrum should also oscillate, which should be detectable by using nano-probes placed on the surface of the 3D TI near the Josephson junction, as illustrated in Fig. 1.

\begin{figure}
\includegraphics[width=0.3 \linewidth]{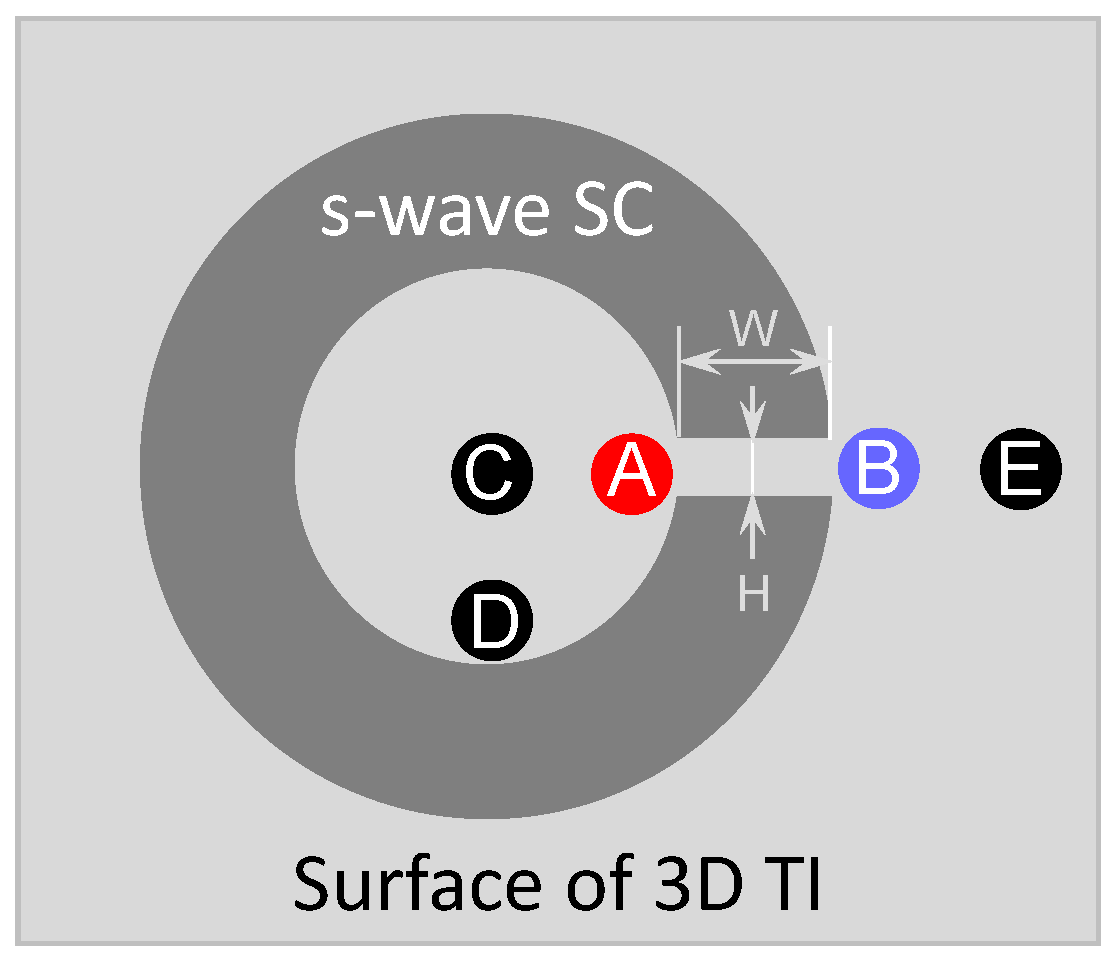}
\caption{\label{fig:fig1} {(color online) Illustration of the rf-SQUID used in this experiment. It contains a superconducting ring with a gap on the surface of a 3D topological insulator (TI). The proximity-effect-induced superconductivity on the TI surface at the gap mediates the Josephson coupling. The dots marked by A, B, C, D and E indicate the places where nano-electrodes are attached to the TI surface for contact resistance measurement.}}
\end{figure}

\begin{figure*}
\includegraphics[width=0.95 \linewidth]{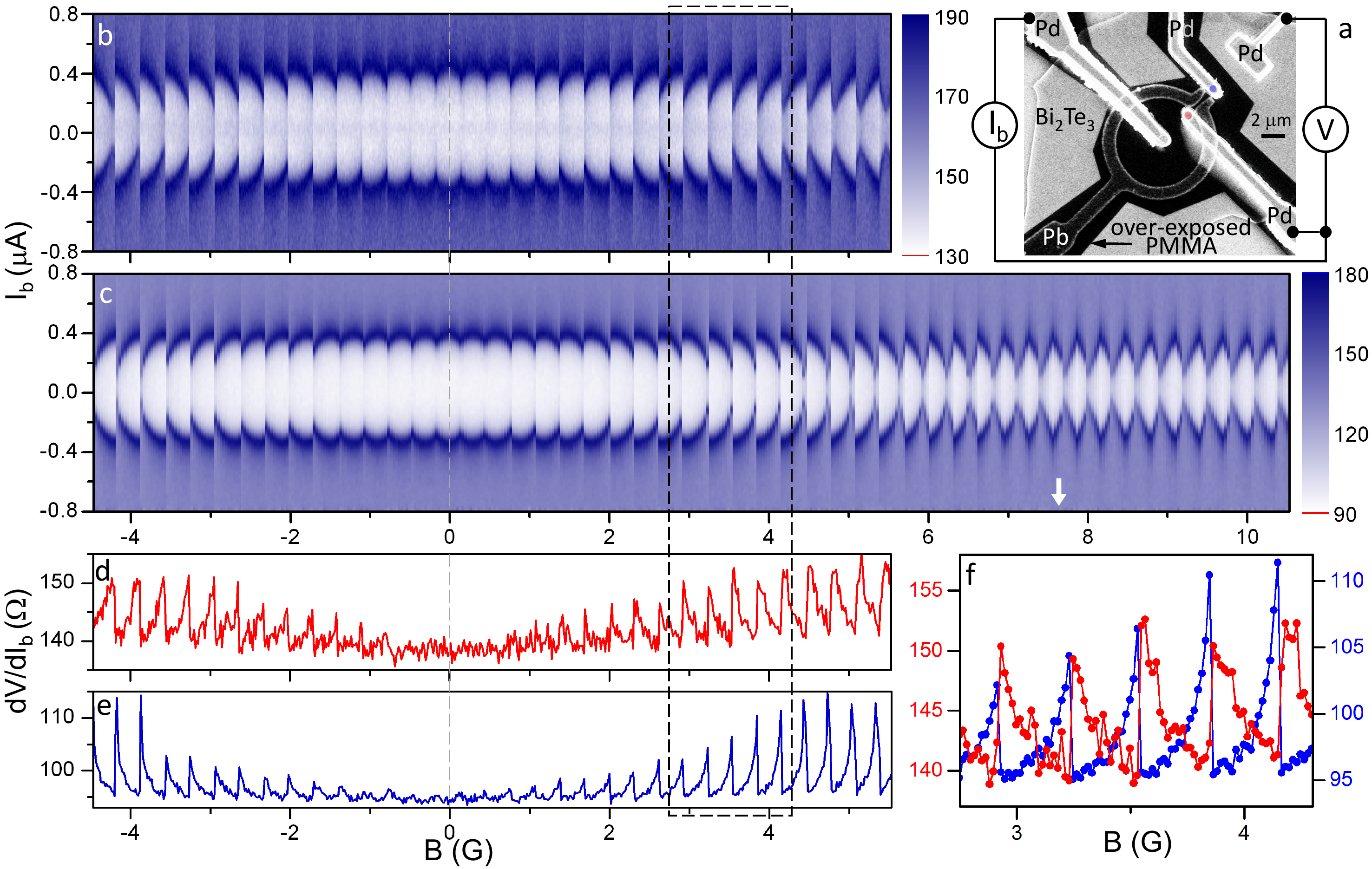}
\caption{\label{fig:fig2} {(color online) Contact resistance $dV/dI_b$ between Pd electrodes and Bi$_2$Te$_3$, measured on one typical device near the inner and outer ends of the Josephson junction at $T$=10 mK. \textbf{a} SEM image of the device and the illustration of a three-terminal configuration for contact resistance measurement. The black area is covered with over-exposed PMMA which prevents the normal-metal Pd electrodes to contact with Bi$_2$Te$_3$ except through small windows at positions A (colored in red), B (colored in blue) and C. \textbf{b} and \textbf{c} 2D plots of $dV/dI_b$ as functions of magnetic field and bias current, measured at positions A and B, respectively. The arrow in c indicates the magnetic field at which the Fraunhofer envelope is expected to reach its first minimum. The color scales are in the unit of $\Omega$. \textbf{d} and \textbf{e} Line cuts of the data in b and c at $I_b$=0, respectively. \textbf{f} The line cuts within the dashed window in d and e are plotted together.}}
\end{figure*}

We have fabricated about twenty such devices by depositing Pb rings on exfoliated Bi$_2$Te$_3$ flakes of $\sim$100 nm in thickness, and measured the contact resistance between normal-metal Pd nano-electrodes and Bi$_2$Te$_3$ at positions A, B, C and D as marked in Fig. 1, by using standard three-terminal measurement configuration and lock-in amplifier technique at cryogenic temperatures of dilution refrigerators. We found that the contact resistance at positions A and B oscillates and even jumps abruptly with varying magnetic flux enclosed by the Pb ring, whereas no oscillation was found at positions C and D (data are shown in the supplementary materials).

Shown in Fig. 2 are typical results obtained on one of the devices. Figure 2a shows the SEM image of the device. The inner and outer diameter of the Pb ring are $d_{in}=7.8$ $\mu$m and $d_{out}=10.2$ $\mu$m, respectively. The size of the Josephson junction is 2.2 $\mu$m$\times$200 nm. Two Pd electrodes were fabricated and connected to the Bi$_2$Te$_3$ surface at positions A and B about 100 nm away from the two ends of the junction, through two windows of 600 nm in diameter each on over-exposed PMMA mask. Another Pd electrode was similarly connected to the center of the ring (position C).

Figures 2b and 2c are 2D plots of the contact resistance $dV/dI_b$ measured at positions A and B, respectively, as functions of magnetic field and bias current. The $dV/dI_b$ oscillates at a period of $\Delta B=0.32\pm0.01$ G, corresponding to an effective area of $S_{eff}=\phi_0 /\Delta B=62.5$ $\mu$m$^2$ (where $\phi_0=h/2e$ is the flux quantum, $h$ is the Planck's constant, and $e$ the electron charge). This estimated area is in good agreement with the geometric area of the ring after considering flux compression: $\pi d_{in} d_{out}/4=62.45$ $\mu$m$^2$.

The most peculiar feature in the 2D plots is that the low-resistance area has a semilunar shape within many oscillation periods, as if an originally 2$\phi_0$-period (i.e., 4$\pi$-period) oscillation is truncated when the flux in the ring reaches odd multiples of half flux quantum. In addition, the semilunar shapes measured at positions A and B take opposite orientations. Within the 0$^{\rm th}$ envelopes of the Fraunhofer-like pattern ($|B|\lesssim 7.7$ G), the ones measured at position A are centered toward the origin, whereas the ones measured at position B are centered away from the origin. The semilunar shapes reverse their orientation after entered into the 1$^{\rm st}$ envelope of the Fraunhofer-like pattern ($|B|\gtrsim 7.7$ G). We note that the characteristic field of 7.7 G corresponds to an area of 2.2 $\mu$m$\times$1.2 $\mu$m, in agreement with the effective area of the junction after taking into account of flux compression (explained in the supplementary materials).

Shown in Figs. 2d and 2e are the line cuts in Figs. 2b and 2c along $I_b=0$, respectively. With varying magnetic flux, the two sets of data measured at positions A and B vary in a complementarily correlated manner, and swap their status abruptly when the flux in the ring reaches every odd multiples of half flux quantum. Then the processes start over again. The two sets of data in the dashed window in Figs. 2d and 2e are plotted together in Fig. 2f, showing that there are two 4$\pi$-period trends differed by a phase shift of 2$\pi$ (i.e., $\phi_0$).

It has been shown previously that superconductivity can be induced across the Pb-Bi$_2$Te$_3$ interface, spreading along the surface in Bi$_2$Te$_3$ up to a distance of microns at low temperatures \cite{QuFM, Yang_Fan2}. The measured contact-resistance oscillations at positions A and B therefore reflect the variation of the superconducting gap in Bi$_2$Te$_3$ at the ends of the Pb-Bi$_2$Te$_3$-Pb junction. The latter is further controlled by the EPR of the junction.

The contact resistance of an S-N interface is in general described by the Blonder-Tinkham-Klapwijk (BTK) theory \cite{BTK}. When the interfacial barrier is high such that the electron transport across the interface is in the tunneling limit, the contact resistance mainly reflects the tunneling density of states of the superconducting side. When the interfacial barrier is low, Andreev reflection becomes a dominant process, resulting in a grossly enhanced conductance (dirty case) or resonantly enhanced conductance peak(s) (clean case) within the gap energy. In our experiment both the high- and low-interfacial barrier cases appeared. In the latter case the low-resistance state in the 2D plot could even have a boundary mimicking a critical supercurrent, as shown in Figs. 2b and 2c, as if proximity-effect-induced superconductivity has developed across the Bi$_2$Te$_3$-Pd interface. This boundary is related to the excess current \cite{Josephson} at the S-N interface due to Andreev reflection.

For the mechanism of jumping, let us firstly point out that it is {\it not} related to the jumping of the 2$\pi$-period screening supercurrent $I_{s,2\pi}$ in the ring --- a conventional jumping mechanism which is well known to occur when the SQUID's screening parameter $\beta_e=2\pi LI_c/\phi_0>1$ (where $L$ is the inductance of the ring and $I_c$ is the critical supercurrent of the junction \cite{Josephson}). In fact, for most of the devices investigated in our experiment, unless otherwise mentioned, their estimated $\beta_e$ was safely below 1 at the base temperature (a list of $\beta_e$ of our devices can be found in Table S1 of the supplementary materials), and was further reduced at elevated temperatures due to the decrease of critical supercurrent.
For devices with $\beta_e<1$, it is known that their $I_{s,2\pi}$ should not jump at any flux. More specifically, $I_{s,2\pi}$ should cross zero smoothly at odd multiples of half flux quantum. Therefore, we conclude that the $dV/dI_b$ jumping we observed is not caused by the jumping of conventional 2$\pi$-period screening supercurrent in the ring. This conclusion is further supported by a control experiment performed on a device with $\beta_e=8.9$ (the details are presented in the supplementary materials).

To test whether the $dV/dI_b$ jumping is unique to the induced topological superconductivity on TI, we have performed another control experiment on graphite-based devices, in which the induced superconductivity in the junction area is supposed to be topologically trivial. The results are presented in the supplementary materials as well.
It is known that the $I_{s,2\pi}$ there in the ring must oscillate in its full strength in order to compensate the flux change. However, the contact resistance between Pd and graphite responses very smoothly, showing no sign of jumping (see Fig. S7-2 in the supplementary materials).
It indicates that the jumping in $dV/dI_b$ is a unique feature for the devices constructed on Bi$_2$Te$_3$.

The abrupt jumping at odd multiples of half flux quantum reflects the existence of a 2$\pi$-period but fully skewed current-phase relation (CPR): $I_{s}=I_{c}{\rm sin}(\varphi/2)$ for $\varphi\in [-\pi, \pi]$. Theoretically, such a CPR could occur in any Josephson junctions made of conventional materials, as long as the quasiparticle transport in the junction is fully transparent (i.e., with a transmission coefficient $D=1$) so that the minigap caused by Andreev reflections has a 2$\pi$-period form of \cite{Josephson}: $\Delta\propto |\cos (\varphi/2)|$. In practical case, however, the jump will be rounded up because $D$ will always be reduced from 1 at finite temperatures and/or due to the existence of imperfections. For example, even for a Josephson junction using single atomic layer of graphene as the barrier and reached a transmission of as high as $D=0.99$, its jumping in CPR is already rounded up significantly \cite{HJLee}. For our junctions $D$ is much lower. Previous study on interfacial conductance \cite{Yang_Fan} shows that the barrier strength $Z$ of Bi$_2$Se$_3$-Sn interface fabricated with the same technique is around 0.6. It corresponds to a transmission coefficient of $D=1/(1+Z^2)=0.74$ -- far from being fully transparent. Therefore, the appearance of fully skewed and truncated patterns in $dV/dI_b$, not only at the base temperature but also at elevated temperatures (e.g., see Figs. S6-1 and S6-2 in the supplementary materials), is rather unusual.

A fully skewed CPR could arise from a topologically protected mechanism involving two branches of EPRs. According to the theories \cite{Fu_liang_2008,Fu_Kane_2009ring,Lutchyn_Sau_Das_Sarma_2010ring,Beenakker_2013ring,Benjamin_Zhang_Kane_2014ring}, for Josephson junctions constructed on TI surface, there are two branches of 4$\pi$-period EPRs when the number of quasiparticle in the junction is a good quantum number. Depending on odd or even number of quasiparticles in the system, the two branches belong to two different macroscopic quantum states with odd or even parity. If the system traces the lowest-energy branch at their crossing points, then a 2$\pi$-period and fully skewed CPR will be yielded. Different from the trivial mechanism containing only one branch of EPR as discussed in the previous paragraph, branch-switching between the two EPRs changes the parity of the system, so that it must be accompanied with adding or removing one quasiparticle from the system known as quasiparticle poisoning \cite{QP_poisoning_Loss,QP_poisoning_Beenakker_2013,Yacoby_QPP}. Besides the yielded fully skewed CPR, the happening of quasiparticle poisoning will also lead to a $\pi$ phase shift, which can be regarded as a smoking-gun evidence whether parity-switching really happens or not in the system. For our device, the number of quasiparticles is unfixed because the junction is connected to the quasiparticle bath of the surroundings. Therefore, quasiparticle poisoning will happen unavoidably when the two branches cross with each other at odd multiples of half flux quantum, if the two-branch mechanism really plays a role.

\begin{figure*}
\includegraphics[width=0.9 \linewidth]{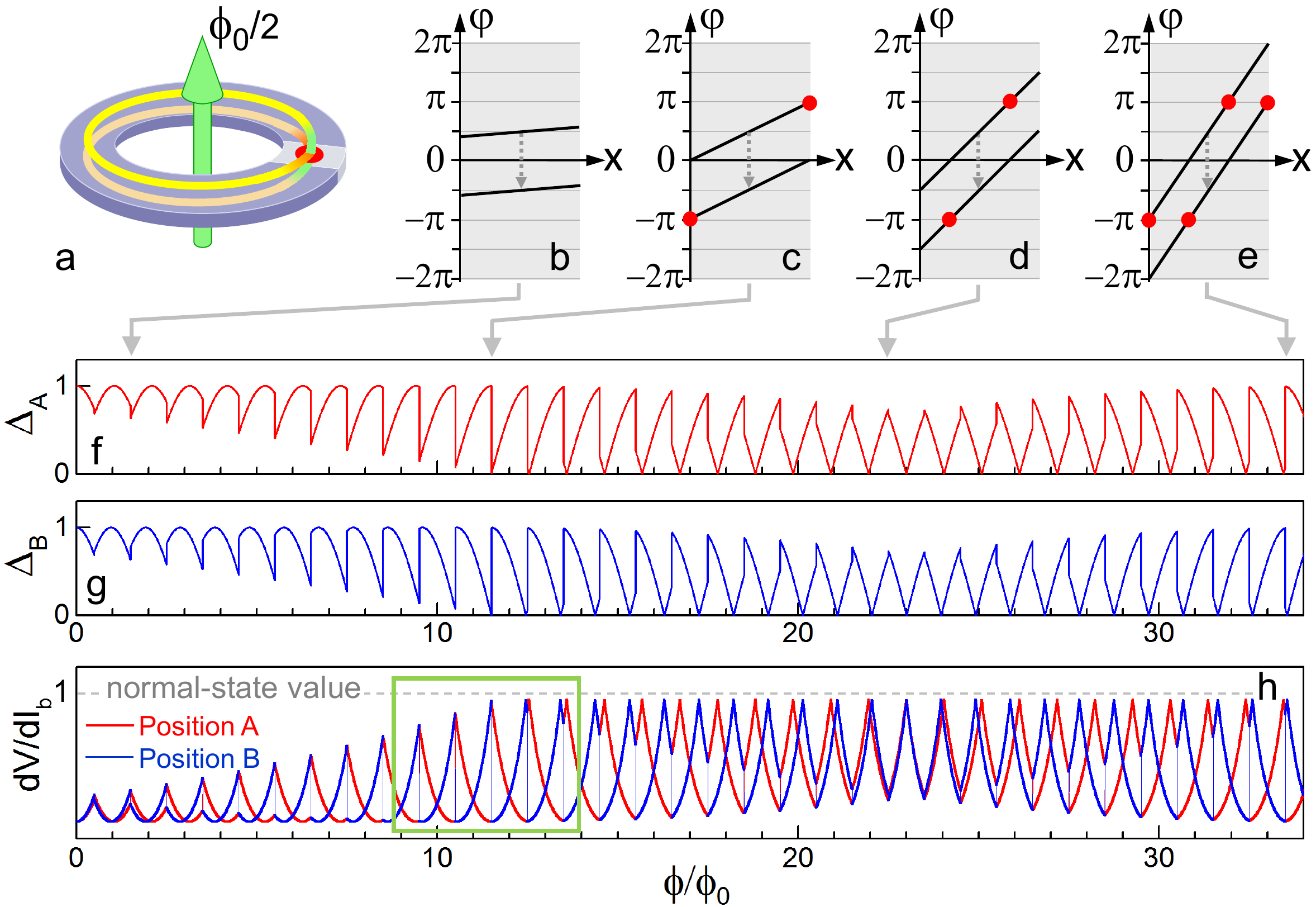}
\caption{\label{fig:fig3} {(color online) \textbf{a} Formation of a $1/2$ fractional mode of Cooper pairs like a M\"{o}bius strip on the ring, as reflected by the half phase-to-flux ratio of $\alpha=1/2$. \textbf{b} to \textbf{e} Distribution and evolution with flux of the local phase difference in the junction. The left/right edge of each plot corresponds to position A/B of the junction as defined in Fig. 1. The two lines in each plot represent the states before and after quasiparticle poisoning. The red dots represent the places where the minigap is closed and Majorana fermion states are believed to occur \cite{Fu_SJJ_2013}. \textbf{f} and \textbf{g} Calculated minigap at positions A and B, respectively, with lineshapes mimicking the boundaries of the low-resistance state of $dV/dI_b$ shown in Figs. 2b and 2c. \textbf{h} Deduced $dV/dI_b$ at $I_b=0$, mimicking the measured line cuts shown in Figs. 2d, 2e and 2f.}}
\end{figure*}

In the following let us present a detailed interpretation for the boundaries and the envelope of the semilunar shapes shown in Figs. 2b and 2c, as well as for the line cuts shown in Figs. 2d, 2e and 2f. It is known that the boundary of the low-resistance state represents a characteristic excess current $I_e$ caused by Andreev reflection at the Bi$_2$Te$_3$-Pd interface \cite{BTK,Josephson}: $I_e\propto (\Delta/eR)\tanh(V/2k_BT)$, where $R$ is the normal-state resistance of the interfacial junction, and $\Delta$ is the local minigap in Bi$_2$Te$_3$. For the surface helical electrons with protected full transparency, their Andreev reflections in the Pb-Bi$_2$Te$_3$-Pb junction yield a local minigap of the form: $\Delta (x)\propto |\cos (\varphi (x)/2)|$. And the local phase difference $\varphi (x)$ is further controlled by the flux $\phi$ in the ring and in the junction via:
\begin{equation}
\varphi (B, x)=2\pi\alpha\frac{\phi}{\phi_0}+2\pi\beta\frac{xH^*B}{\phi_0}\pm\pi\hspace {0.05 cm}{\rm int}(\frac{\phi}{\phi_0}+\frac{1}{2})
\end{equation}
where $\phi=BS_{eff}$, $x$ is defined from $-W/2$ (position A) to $W/2$ (position B), $W=2.2$ $\mu$m is the width of the junction, and $H^*=1.2$ $\mu$m is the effective length of the junction after considering flux compression (explained in the supplementary materials). The first term in Eq. (1) represents the phase difference generated by the flux in the ring. The second term is the phase difference generated by the flux in the junction. It determines the slope of the the curves in Figs. 3b to 3e. The third term represents an additional $\pi$ phase shift caused by quasiparticle poisoning at every odd multiples of half flux quantum.

The calculated $\Delta$ at positions A and B are shown in Figs. 3f and 3g, respectively. The results pertinently describe the boundaries of the low-resistance state shown in Figs. 2b and 2c, with no fitting parameter except for taking $\alpha=1/2$ and $\beta=1$. A more detailed fitting can be found in the supplementary materials. Furthermore, by tentatively taking $dV/dI_b\propto e^{-\Delta/k_BT}$ (where $k_B$ is the Boltzmann constant), the $dV/dI_b$ at $I_b=0$ can also be mapped (Fig. 3h). The results mimic the line cuts shown in Figs. 2d, 2e and 2f. In particular, the curves within the green window in Fig. 3h correspond to the ones shown in Fig. 2f. We note that a rigorous mapping between $dV/dI_b$ and $\Delta$ would require numerical calculations within the BTK theory \cite{BTK}.

For ordinary Cooper pairs the phase-to-flux ratio must be $\alpha=\beta=1$. The experimental finding of $\alpha=1/2$ reflects the formation of $1/2$ fractional modes of Cooper pairs on the ring as represented by the M\"{o}bius strip in Fig. 3a. Since the phase generated by the flux can in general be expressed as \cite{Wilczek} $\varphi=2\pi\phi/\phi^*_0$, the half phase-to-flux ratio corresponds to an effective flux quantum of $\phi^*_0=h/e^*=2\phi_0$, where $e^*=e$ is the effective charge of fractionalized Cooper pairs that passes through the junction, mimicking in a superconducting Kitaev chain \cite{Kitaev2001}. It is the appearance of $\alpha=1/2$, not the full-transparency-caused 1/2 factor in the minigap formula, that provides the evidence for the existence of the long-sought 4$\pi$ periodicity, despite that the measured periods are 2$\pi$-periodic after being truncated by quasiparticle poisoning.

Contrary to the global mode on the ring which passes through the TI surface once to form the 1/2 fractional mode, the local mode surrounding the Josephson vortex in the junction passes through the TI surface twice. We believe that this guarantees the formation of integer modes in the junction, so that the phase-to-flux ratio for the second term in Eq. (1) is $\beta=1$. More theoretical study would be needed to discuss this experimental finding.

According to Potter and Fu \cite{Fu_SJJ_2013}, the red dots in Figs. 3c to 3e represent the positions where the minigap is closed so that Majoranas are expected. The positions of Majoranas can be manipulated by varying the flux, and swapped by quasiparticle poisoning when the two $4\pi$-period EPRs cross with each other at odd multiples of half flux quantum. Figures 3c and 3e further explain that, when the flux in the junction area reaches half flux quantum (i.e., $\sim 11.5\phi_0$ in Fig. 3), the Majoranas move to the ends of the junction, so that the contact resistance at position A/B experiences full gap-closing, together with the sharpest jumping before and after position swapping.

To conclude, we have observed clear evidence for the existence of $1/2$ fractional modes and the occurrence of quasiparticle poisoning. These are necessary signatures accompanied with the formation of Majorana fermion states. Our study demonstrates that superconducting devices constructed on the surface of 3D TIs provide a promising platform for braiding Majorana fermions in two dimensions and building topological quantum computers in the future.

\vspace {0.5 cm}

\noindent\textbf{Acknowledgments} We would like to thank S. Y. Han, C. Beenakker, R. Du, P. A. Lee, K. T. Law, S. P. Zhao, Z. Fang, X. Dai, T. Xiang, G. M. Zhang, X. Hu and L. Yu for fruitful discussions. This work was supported by the National Basic Research Program of China from the MOST under the grant No. 2009CB929101 and 2011CB921702, by the NSFC under grant No. 91221203, 11174340, 11174357, 91421303 and 11527806, and by the Strategic Priority Research Program B of the Chinese Academy of Sciences under the grant No. XDB07010100.

\vspace {0.5 cm}

\noindent\textbf{Author contributions} Y.P., J.S., F.Q. and J.W performed the experiment. J.F., G.L., Z.J., X.J. and C.Y. provided experimental assistance. Y.P. and L.L. wrote the manuscript. All the authors participated in the discussion.

\vspace {0.5 cm}

\noindent\textbf{Additional information} Supplementary information is available in the online version of the paper. Reprints and permissions information is available online at www.nature.com/reprints. The authors declare no competing financial interests.



\section*{Supplementary material}

# Supplementary Materials for
## "Majorana fermions in a superconducting Mobius strip"

Yuan Pang[1],*, Jie Shen[1],*, Fanming Qu[1],†, Zhaozheng Lyu[1], Junhua Wang[1], Junya Feng[1], Jie Fan[1], Guangtong Liu[1], Zhongqing Ji[1], Xiunian Jing[1,2], Changli Yang[1,2], Qingfeng Sun[2,3], X. C. Xie[2,3], Liang Fu[4] and Li Lu[1,2,‡]

*1 Beijing National Laboratory for Condensed Matter Physics, Institute of Physics, Chinese Academy of Sciences, Beijing 100190, People's Republic of China*

*2 Collaborative Innovation Center of Quantum Matter, Beijing 100871, People's Republic of China*

*3 International Center for Quantum Materials, Peking University, Beijing 100871, People's Republic of China*

*4 Department of Physics, Massachusetts Institute of Technology, Cambridge, Massachusetts 02139, USA*

* These authors contributed equally to this work.

† Present address: QuTech and Kavli Institute of Nanoscience, Delft University of Technology, 2600 GA Delft, The Netherlands

‡ Corresponding author: lilu@iphy.ac.cn

## Contents

## 1. Contact resistance between Pd and Bi₂Te₃ at other positions of the rf-SQUID

We have fabricated several devices which contain Pd electrodes not only at positions A and B, but also at positions C and D defined in Fig. 1 of the main manuscript. Figure S1a shows the SEM image of one of such devices (device #S1). The 2D plot of contact resistance $dV/dI_b$ measured at positions A, C, and D are shown in Figs. S1 b, c, and d, respectively. Only at positions near the line Josephson junction, e.g., at position A, $dV/dI_b$ oscillations were observed with varying magnetic flux in the Pb ring. No resistance oscillations were found at positions away from the line Josephson junction, namely positions C and D, no matter their positions are close to the superconducting ring or not. The results indicate that the oscillations we observed are only related to the status of the Josephson junction in the rf-SQUID loop.

Although there is no oscillation, the $dV/dI_b$ data measured at position D still show a low-resistance state within a characteristic gap energy, as shown in Fig. S1d. This low-resistance state is caused by the induced superconductivity in Bi₂Te₃ in the vicinity of the Pb ring. In the main manuscript, we attribute the low-resistance state to Andreev reflection and/or proximity superconductivity at the secondary interface (i.e., between Pd and Bi₂Te₃).

In addition to the low-resistance state, there is also a zero-bias resistance peak (ZBRP) in Fig. S1d. This ZBRP is a common feature of the secondary interface, regardless there are oscillations or not --- the ZBRP superimposes with the $dV/dI_b$ oscillations at positions A and B when the contact resistance is relatively high (see Fig. S2-1b, also Figs. S2-2a and S2-2b). This ZBRP at the secondary interface is directly related to the zero-bias conductance peak (ZBCP). The latter is reproducibly observed at the primary interface [1]. It is presumably related to the gap-closing in Bi₂Te₃ at the Fermi energy.

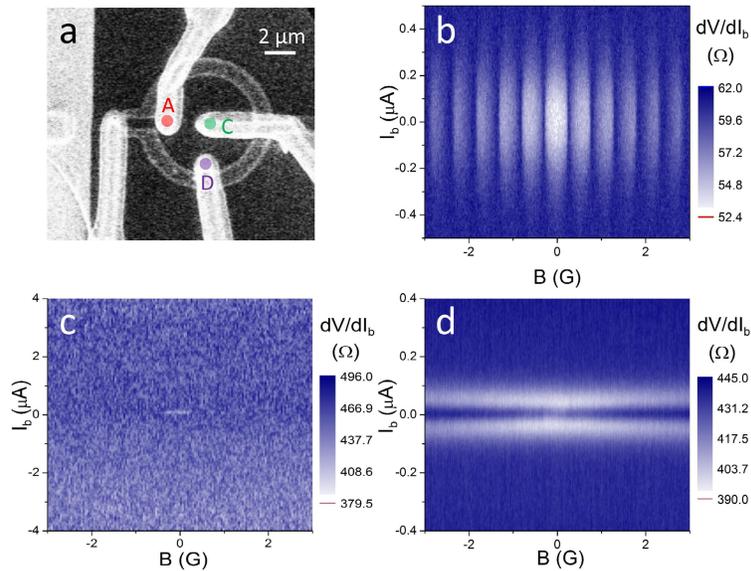

**FIG. S1 |** (**a**) SEM image of device #S1. (**b**), (**c**) and (**d**) The $dV/dI_b$ measured at positions A, C and D, respectively, as functions of magnetic field and bias current at T=10 mK.

2

## 2. Data obtained on Pd electrodes with relatively high contact resistance

As mentioned in the main manuscript, for the data taken by electrodes with relatively low contact resistance, there is a clear boundary in the 2D plots of $dV/dI_b$ mimicking a critical supercurrent for the low-resistance state, indicating that proximity superconductivity has developed across the S-N interface.

We have also investigated a number of devices with relatively higher contact resistance between $Bi_2Te_3$ and Pd, but still in the regime of conductance enhancement within the gap. We found that their low-resistance state no longer has a clear semilunar boundary anymore. The absence of such a boundary indicates that proximity superconductivity is not yet developed across the $Bi_2Te_3$-Pd interface, only conductance enhancement via Andreev reflection plays a role there. Nevertheless, semilunar-like shapes still emerge in the 2D plots.

In Fig. S2-1 we show the data obtained on device #S2 which has electrodes with both relatively high and low contact resistance. The data in Fig. S2-1c, where the low-resistance area has a clear boundary, indicate that the S-N interface at position B has become superconducting. This argument is further supported by the fact that the characteristic critical supercurrent increases with decreasing temperature, as will be shown later in Figs. S6 f, g, h, I, and j. On the other hand, for the data shown in Fig. S2-1b, the low-resistance area does not have a clear boundary. It indicates that proximity superconductivity is not yet well developed across the S-N interface at position A, only conductance enhancement via Andreev reflection occurs. This argument is again supported by the temperature variation shown in Figs. S6 a, b, c, e, and f. With decreasing temperature, the enhanced conductance peak gets narrower and narrower against both bias current and bias voltage, exhibiting a resonant nature.

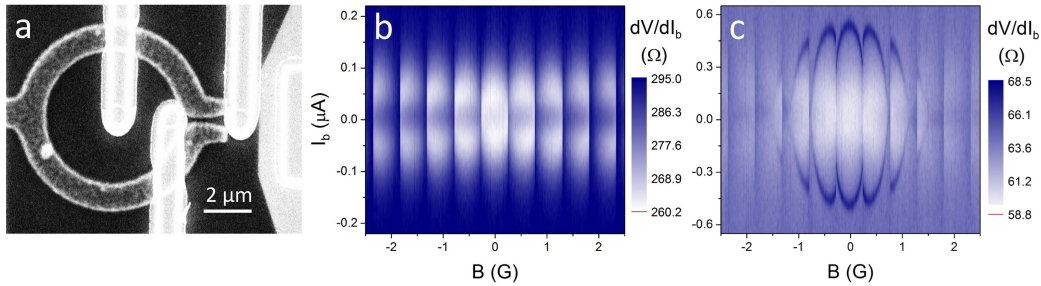

**FIG. S2-1 |** The contact resistance $dV/dI_b$ of device #S2 measured at 10 mK. (**a**) An SEM image of the device. (**b**) 2D plot of $dV/dI_b$ measured at position A. The oscillating low-resistance region does not have a clear boundary. (**c**) 2D plot of $dV/dI_b$ measured at position B. The low resistance regions have a clear boundary.

Regardless of relatively high or low contact resistance, the jumping of $dV/dI_b$ with magnetic flux can always be seen at positions A and B, both on the line cuts of the 2D plot and through the hyperbolic-like and concentric-like semilunar shapes in the 2D plot. For electrodes with high contact resistance, the jumps on the line cuts appear to be even sharper. When the magnetic flux is slightly away from zero, the contact resistance periodically oscillates to its normal-state value, reflecting that gap-closing happens in $Bi_2Te_3$ at odd multiples of half flux quantum.

In Fig. S2-2 we show more data obtained on device #S3 whose both electrodes at positions A and B have relatively high contact resistance. There are sharp jumpings to the normal-state resistance (namely gap-closing) on the line cuts.

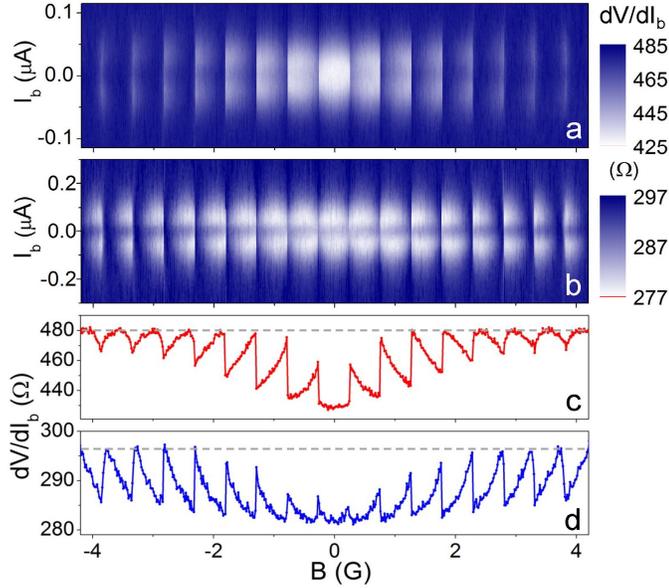

**FIG. S2-2 |** Oscillatory patterns of $dV/dI_b$ obtained on another device (device #S3) with relatively high contact resistance. The measurement was performed at T=10 mK. (**a**) and (**b**) $dV/dI_b$ measured at positions A and B, respectively. Although without a clear boundary, hyperbolic-like and concentric-like semilunar shapes still emerge. (**c**) and (**d**) Line cuts in (a) and (b) at $I_b = 0$, respectively. Gap-closing in $Bi_2Te_3$ happens at odd multiples of half flux quantum, as that the $dV/dI_b$ hits the normal-state values (the dashed lines) of the contact resistance at these flux.

## 3. Discussions on $\beta_e$ and a list of $\beta_e$ for the devices investigated

In applied magnetic field, screening supercurrents will be induced in our device, circulating along the Pb ring and around the Josephson junction area. When the critical supercurrent of the junction is large and/or the inductance of the loop is large, the induced supercurrent in the ring can significantly influence the total magnetic flux in the ring, trying to make it quantized.

The total magnetic flux in the loop is [2]:

$$\Phi = \Phi_e - (\beta_e \phi_0 / 2\pi) \sin(2\pi\Phi/\phi_0) \tag{1}$$

where $\Phi_e$ is external flux exerted by the applied magnetic field, $\phi_0$=h/2e is flux quanta, $\beta_e$=2πL$I_c$/$\phi_0$ is the SQUID screening parameter, L is the inductance of the ring, and $I_c$ is the critical supercurrent of the junction in the superconducting loop.

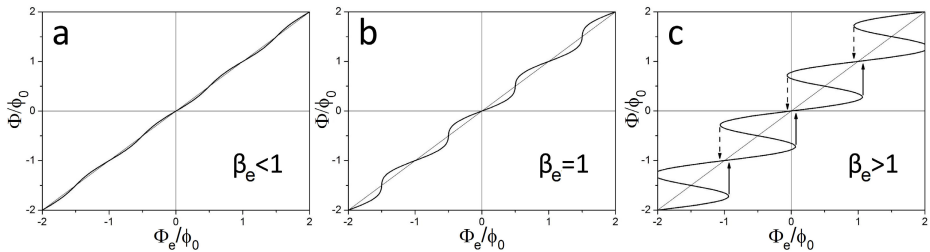

**FIG. S3 |** The total magnetic flux $\Phi$ vs. the external magnetic flux $\Phi_e$ for SQUID loops with different SQUID screening parameter $\beta_e$. (**a**) $\beta_e$<1. (**b**) $\beta_e$=1. (**c**) $\beta_e$>1. For the case (c), flux jumping, 2π-period circulating supercurrent jumping, and hysteresis between opposite directions of field sweeping occur.

The dependence of the total flux $\Phi$ on the external flux $\Phi_e$ shows two different kinds of behavior according to the value of $\beta_e$. For $\beta_e \leq 1$, $\Phi$ is single valued and increases monotonically with increasing $\Phi_e$ (Fig. S3 a and b). For $\beta_e > 1$, $\Phi$ is multivalued on $\Phi_e$, only the segments of the curve with positive slope are traced in field sweeping (Fig. S3c). There is hysteresis between opposite directions of field sweeping. The total flux of the ring and the circulating supercurrent in the ring jump at positions determined by the specific value of $\beta_e$, not necessary at odd multiples of half flux quanta.

To determine the $\beta_e$ of our rf-SQUID, we need to know the critical supercurrent of the Josephson junction in the SQUID loop which cannot be directly measured because the junction is short-circuited with the superconducting loop. We thus have to fabricate a single Josephson junction of the same size and on the same flake of $Bi_2Te_3$ or graphite, then using the critical supercurrent of the single Josephson junction to estimate the $\beta_e$ of the SQUID. Some of the results are summarized in Table S1.

**Table S1 |** The $\beta_e$ of some of the devices we investigated

| Device number | Device presented in the main manuscript | #S2 | #S3 | #S4 | #S5 |
|---|---|---|---|---|---|
| Geometric parameters of the Pb rings and the junctions | $D_{in}$=7.8 μm $D_{out}$=10.2 μm W=2.2 μm | $D_{in}$=5.7 μm $D_{out}$=8 μm W=2.2 μm | $D_{in}$=5.7 μm $D_{out}$=8.1 μm W=3.2 μm | $D_{in}$=4 μm $D_{out}$=5.8 μm W=2.2 μm | $D_{in}$=18 μm $D_{out}$=20 μm W=5.2 μm |
| Inductance L | 9.9 pH | 5.7 pH | 5.8 pH | 2.9 pH | 44.8 pH |
| Estimated Ic of the junction | ~ 27 μA | ~ 27 μA | ~ 39 μA | ~ 27 μA | ~ 63 μA |
| Estimated $\beta_e$ | 0.84@10 mK | 0.48@10 mK | 0.71@10 mK | 0.25@20 mK | 8.9@20 mK |

Notes:  $D_{in}$ and $D_{out}$ are the inner and outer diameters of the Pb ring, respectively.
W is the width of the Josephson junction. The length H of the junctions is the same for all devices, being around 200-300 nm.

For most of the devices we investigated, $\beta_e$ is smaller than one, so that the devices are in the non-hysteretic regime. For example, the data taken on device #S4 are shown in Fig. S4c. No hysteresis is seen between the black and red lines taken in opposite directions of field sweeping.

## 4. More data obtained on devices with $\beta_e < 1$

Shown in Fig. S4 are the data obtained on another device #S4 whose estimated SQUID screening parameter is $\beta_e = 0.25$. It has a slightly smaller diameter and also a slightly higher contact resistance than that of the device shown in Fig. 2 of the main manuscript. The temperature and field dependencies of the data clearly indicate that it is the resistance dip (i.e., the conductance peak) near zero bias voltage that oscillates with the magnetic flux in the ring. The energy scale (width) of this dip that oscillating is around 0.2 mV, an order of magnitude larger than that of the ZBRP structure shown in Fig. S1d, indicating that this dip is an Andreev-reflection-induced gross conductance enhancement within the gap of superconducting $Bi_2Te_3$. Proximity superconductivity is not yet developed across the S-N interface, because the low-resistance state in Fig. S4b does not have a clear boundary. The amplitude of resistance oscillation in Fig. S4c corresponds to an conductance enhancement of 0.45 $e^2/h$. Figures S4 e and f show that, with decreasing temperature, the width of the dip gets narrower, and the depth of the dip gets deeper, showing a resonant nature. This indicates again that the low-resistance state is caused by Andreev reflection, not the appearance of a supercurrent.

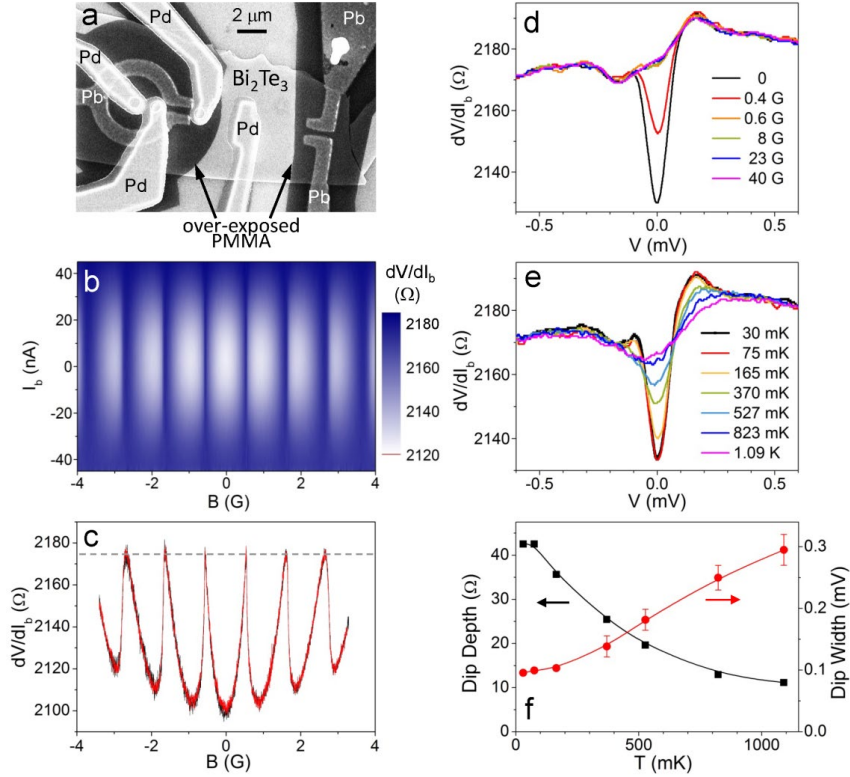

**FIG. S4 |** The results obtained on another device #S4 with $\beta_e \approx 0.25$. (**a**) An scanning electron microscopy image of the device. The single Josephson junction identical to that on the ring was used for estimating the critical supercurrent of the junction. (**b**) 2D plot of $dV/dI_b$ measured at position B (near the end of the junction outside the ring) at T=30 mK. (**c**) $dV/dI_b$ at line cut $I_b=0$, measured in two opposite field-sweeping directions (lines in black and red, respectively), showing no hysteresis. Also, the curves periodically hit the normal-state value represented by the dashed line, indicating gap-closing in $Bi_2Te_3$ at these flux. (**d**) The bias voltage dependence of $dV/dI_b$ measured at position B at T=30 mK and in different magnetic fields. (**e**) The temperature evolution of the bias voltage dependence of $dV/dI_b$ measured at position B at zero field. (**f**) The temperature dependencies of the depth (black squares) and the full width at half depth (red circles) of the dip. The ac excitation current was 20 nA except in (**c**) where it was 2 nA.

## 5. Control experiments (I): data obtained on a devices with $\beta_e > 1$

To demonstrate that our estimation of $\beta_e$ is reliable, we have performed a control experiment on a device with $\beta_e > 1$. In this case, $I_{s,2\pi}$ is expected to jump via the conventional mechanism, at the places which is not necessary located at odd multiples of half flux quantum but rather depending on the specific value of $\beta_e$. And hysteresis in jumpings is expected if the field sweeping direction is reversed.

In Fig. S5 we show the data obtained on device #S5 which has a large loop area and a long junction width (5.2 $\mu$m). Its screening parameter is estimated to be $\beta_e \approx 8.9$. Obvious hysteresis happens when the field sweeping direction is reversed.

We must point out that in the conventional mechanism there should have no oscillation and jumping in the hysteresis region. The jumping observed here in this region is presumably still caused by quasiparticle poisoning, the accumulation of which could even prevent the happening of the conventional jumping. Nevertheless, the position of jumping now is influenced by the screening supercurrent via the conventional mechanism, so that the places of jumping for this device are not located at every odd multiples of half flux quanta, but near the integer multiples of flux quanta. This is understandable. If the conventional screening supercurrent $I_{s,2\pi}$ is large enough in large-$\beta_e$ device, and is able to modify the total magnetic flux in the ring, then the places of jumping caused by the anomalous $I_{s,4\pi}$ supercurrent will be shifted accordingly.

On one hand, the results tell us that the places of jumping in $dV/dI_b$ can be influenced and shifted away from odd multiples of half flux quantum in the $\beta_e > 1$ case. On the other hand, the results confirm that the rest devices we investigated are all in the $\beta_e < 1$ regime, for that their jumpings always take place precisely at odd multiples of half flux quantum, showing no hysteresis in bi-directional field sweeping (e.g., see Fig. S4c). Therefore, we can conclude that the $dV/dI_b$ jumping we observed is not caused by the jumping of the conventional $2\pi$-period screening supercurrent in the ring.

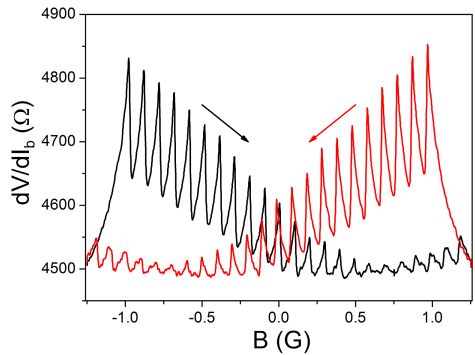

**FIG. S5 |** $dV/dI_b$ curve at position A of device #S5 ($\beta_e \approx 8.9$) as a function of magnetic field, measured in opposite field sweeping directions (illustrated by the arrows). T=20 mK.

## 6. The temperature dependence of the contact resistance oscillation

To further support our argument that the observed jumping in contact resistance is not caused by the jumping of the conventional $2\pi$-period supercurrent $I_{s,2\pi}$ in the ring in the $\beta_e>1$ case, we have performed measurements on several devices not only at the base temperature, but also at elevated temperatures, to further reduce the $I_c$ and thus the $\beta_e$.

Figure S6-1 shows the contact resistance at positions A and B measured on device #S2 ($\beta_e=0.54$ at 10 mK) at several different temperatures. Although $\beta_e$ gets smaller and smaller with increasing temperature, the jumping phenomenon persists and remains to be sharp (Fig. S6-2). Moreover, the places of jumping keep locked at odd multiples of half flux quanta for the data taken at different temperatures. The results convincingly rule out the possibility that the jumping in contact resistance is caused by the jumping of the conventional $2\pi$-period supercurrent $I_{s,2\pi}$ in the $\beta_e>1$ case.

On the other hand, if the occurrence of jumping corresponds to a fully skewed current-phase relation of our devices (i.e., a fully transparent barrier), then it is quite surprised that the jumping keeps sharp at elevated temperatures to $T/T_c\approx0.4 - 0.6$, as shown in Fig. S6-2. The jumping caused by this mechanism usually rounds up with increasing temperature and/or disorders. The results provide strong evidence that the jumping has a non-trivial mechanism.

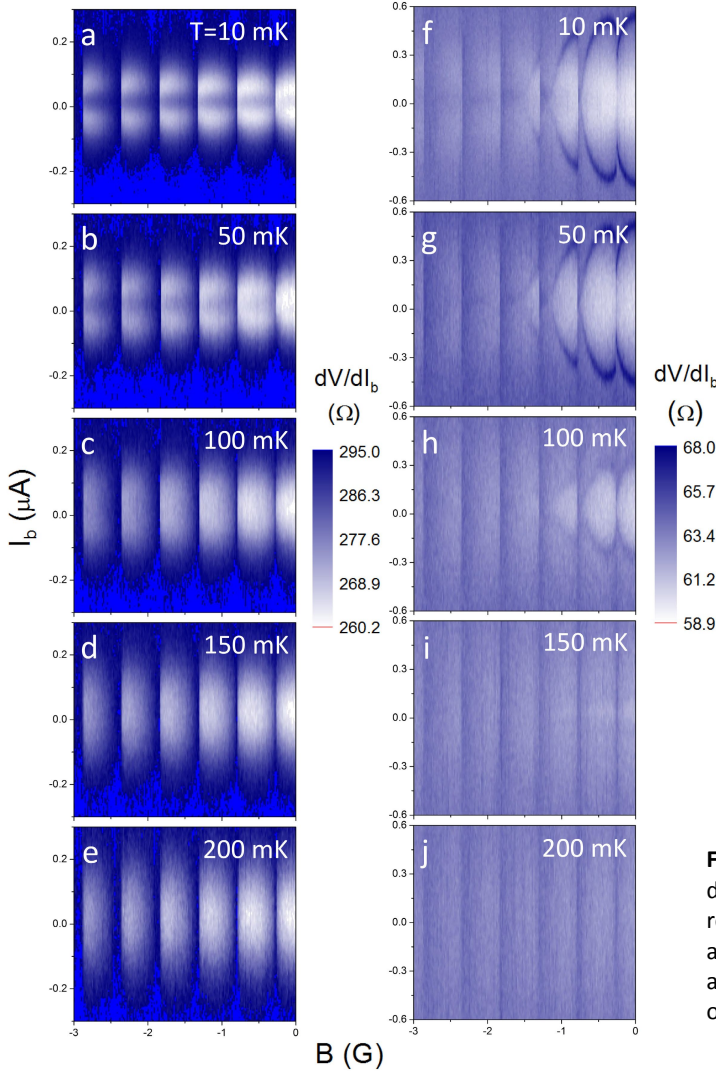

**FIG. S6-1 |** Temperature dependence of the contact resistance dV/dI$_b$ measured at position A (**a**, **b**, **c**, **d**, **e**), and at position B (**f**, **g**, **h**, **i**, **j**) on device #S2.

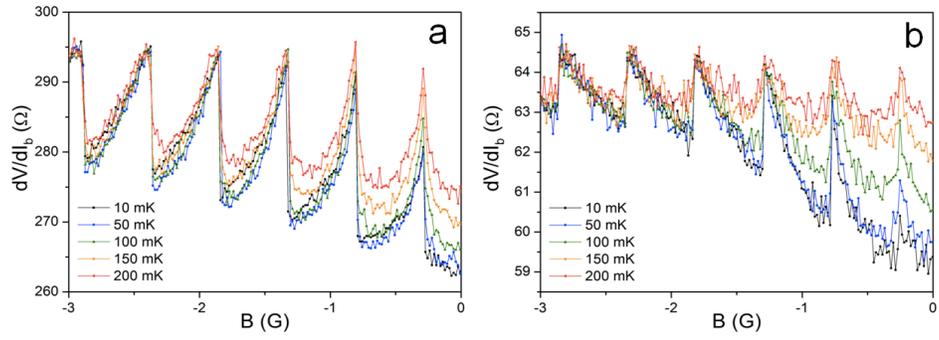

**FIG. S6-2 |** (**a**) Line cuts at $I_b=0$ in Figs. S6-1 (a), (b), (c), (d) and (e), respectively. (**b**) Line cuts at $I_b=0$ in Figs. S6-1 (f), (g), (h), (i) and (j), respectively. The jumpings remain to be sharp at elevated temperatures of $T/T_c \approx 0.4 - 0.6$, where $T_c \approx 0.3$ - $0.5$ K is the critical temperature of proximity-induced superconductivity in $Bi_2Te_3$.

## 7. Control experiments (II): data obtained on graphite-based devices

We have also performed control experiments on graphite-based devices. The results are shown in Figs. S7-1 and S7-2.

Unlike the electron system on the surface of a 3D TI which contains only one type of helical electrons, the electrons in graphene are four-fold degenerated. Therefore, Graphite is known as a topologically-trivial material, with negligible spin-orbit coupling. It is an ideal candidate to be used for performing comparative measurements.

The superconducting proximity effect between Pb and graphite appears to be relatively weak. We therefore have to use Sn to replace Pb. Due to various technique issues, it took us several months to succeed.

In Fig. S7-1 we show the results obtained from one of the graphite-based devices. The contact resistance measurement by Pd electrode at position B revealed a fully developed gap at low temperatures (i.e., with saturated width and amplitude, see Fig. S7-1b), reflecting that the graphite beneath the electrode has become superconducting, owing to the proximity effect from the Sn rf-SQUID. In such a device, the $I_{s,2\pi}$ in the ring must oscillate in its full strength trying to compensate the change of magnetic flux in the ring. However, no noticeable influence on the contact resistance (hence, the gap) was observed during this process, only a Fraunhofer-like pattern of the Josephson junction was seen (Fig. S7-1c).

The Fraunhofer-like pattern shown in Fig. S7-1c is for a high resistance state. It reflects that the gap beneath the Pd electrode in graphite is modulated by the flux in the Josephson junction. A similar Fraunhofer-like pattern for a zero-resistance state was observed on the single Josephson junction in the right side of Fig. S7-1a (data not shown).

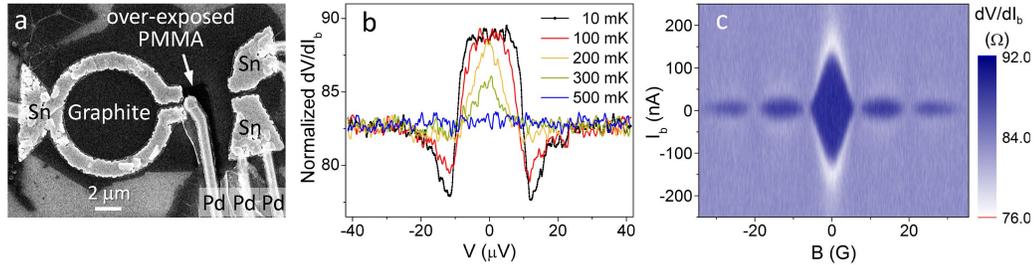

**FIG. S7-1** | A comparative experiment carried out on Sn-graphite device. (**a**) An scanning electron microscopy image of the device. (**b**) The bias voltage dependence of the contact resistance $dV/dI_b$ measured at position B at several different temperatures and in zero magnetic field. (**c**) 2D plot of $dV/dI_b$ measured at position B at T=10 mK, demonstrating a Fraunhofer-pattern-like variation of gap with magnetic flux in the Josephson junction area.

One possibility of observing no contact resistance oscillation but only a Fraunhofer-like pattern is that the probing current $I_b$ (which was applied between the Pd electrode and the Sn ring) was large compared with the critical supercurrent of the junction, so that it disturbed the current flowing in the junction. This could happen because the parameters such as $I_c$ for devices on graphite were not well controlled so far -- the junction on the ring might have a small $I_c$ compared with the applied $I_b$. We therefore investigated more devices.

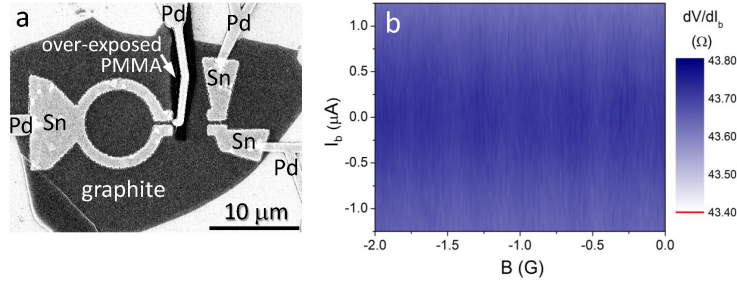

**FIG. S7-2 |** Results obtained on another Sn-graphite device. (**a**) An scanning electron microscopy image of the device. (**b**) 2D plot of $dV/dI_b$ measured at position B at T=10 mK. Note that the range of the color scale is only 1% of the total resistance, much smaller than the ones used for other devices.

In Fig. S7-2 we show the results obtained on another graphite-based device. An oscillatory pattern can be marginally resolved in a period corresponding to the ring area. However, the amplitude of $dV/dI_b$ oscillation is less than 0.3%, about two orders of magnitude weaker than those observed on TI-based devices. And no sign of jumping can be resolved.

Overall, our results indicate that the superconductivity based on topologically-trivial bulk states gives no contribution to the $dV/dI_b$ jumping. Thus, the jumping observed on TI-based rf-SQUIDs is presumably a phenomenon particularly related to the superconducting surface of $Bi_2Te_3$.

In the following we give an explanation for the observed result shown in Fig. S7-2b. Similar to the analysis in the main manuscript, we assume that what the Pd electrode probed is the minigap in graphite at position B near the Josephson junction of the rf-SQUID. And we assume that the quasiparticle transport within the junction area is nearly ballistic so that the minigap can still be expressed as $\Delta \propto |\cos(\varphi/2)|$. Because in graphite the electron states are trivial, the phase-to-flux ratio of the ring is $\alpha=1$. Therefore, $\varphi$ can be expressed as:

$$\varphi(\mathbf{B}, x) = 2\pi\phi/\phi_0 + 2\pi x \mathbf{H}\mathbf{B}/\phi_0$$

where **B** is the magnetic field, x is defined from −W/2 (position A) to W/2 (position B), W is the width and H is the length of the junction.

Figure S7-3 shows the calculated pattern of minigap oscillation at position B (at x=W/2), which is in good agreement with the experimental result shown in Fig. S7-2c as long as the characteristic current in Fig. S7-2c is proportional to the minigap.

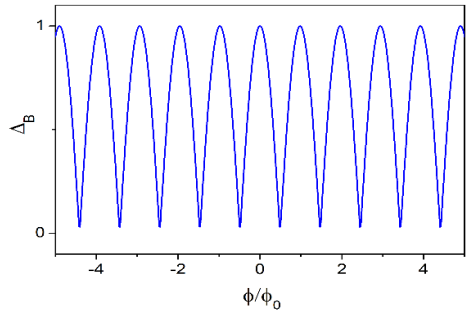

**FIG. S7-3 |** Calculated pattern of minigap oscillation at position B (x=W/2) in graphite-based rf-SQUIDs.

Because electron transport in real graphite devices is not fully transparent, the minigap will never be fully closed, which accounts for the weak oscillation amplitude of $dV/dI_b$.

## 8. Josephson energy profiles in the presence of both 2π-period and 4π-period modes, quasiparticle poisoning and supercurrent reversing

In many phase sensitive experiments, the trivial 2π-period CPR, which would mainly arise from the bulk states, gives considerable contribution. In the presence of both 2π-period and 4π-period modes in a rf-SQUID, the total Josephson energy is the sum of these modes.

For the conventional 2π-period mode, if assuming that the magnetic energy of the ring inductance is small (which is true in our case because of small critical supercurrent and ring area), then the total energy is:

$$E_{J,2\pi} \propto (1/W) \int [1-\cos(2\pi\phi/\phi_0+2\pi\mathbf{B}Hx/\phi_0)]dx \quad \text{(for x=-W/2 to W/2)}$$
$$= -A \sin(\pi\phi'/\phi_0)/(\pi\phi'/\phi_0) \cos(2\pi\phi/\phi_0)$$

where $\phi$ is the magnetic flux in the ring, and $\phi'=\mathbf{B}HW$ is the magnetic flux in the junction, $\mathbf{B}$ is the magnetic field, H is the length and W is the width of the junction.

Similarly, for the two 4π-period modes ($\alpha=1/2$) with a 2π phase shift, their energy can be calculated by using the phase expressed in Eq.(1) of the main manuscript. The results are:

$$E_{J,4\pi} = \pm B \sin(\pi\phi'/\phi_0)/(\pi\phi'/\phi_0) \cos(\pi\phi/\phi_0)$$

where the plus/minus sign represents the odd/even parity branch.

Thus the total energy of the device can be written as:

$$E_J = \sin(\pi\phi'/\phi_0)/(\pi\phi'/\phi_0) [-A \cos(2\pi\phi/\phi_0) \pm B \cos(\pi\phi/\phi_0)]$$

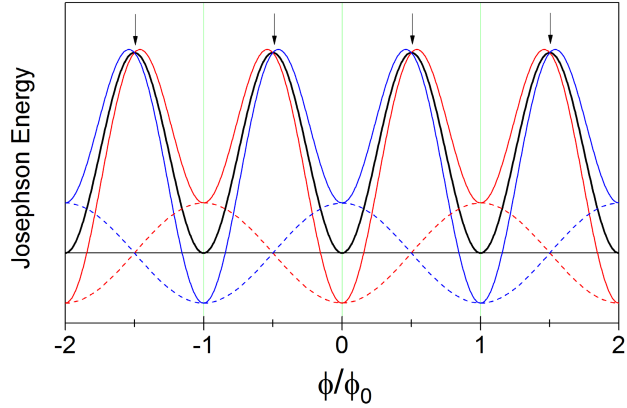

**FIG. S8 |** Energy profiles of a rf-SQUID in the presence of both 2π- (black) and 4π-period modes (dashed red and blue) in the vicinity of $\phi' \approx 0$. The total energy is represented by the solid red and blue curves. The arrows indicate where branch-crossing happens.

In Fig. S8, $E_{J,2\pi}$ is plotted in black line, and $E_{J,4\pi}$ is plotted in dashed red and dashed blue lines for even and odd parity modes, respectively. The total energy of the system is plotted in solid red and solid blue lines, assuming that $B=0.5A$.

It can be seen that the place where the two branches cross with each other is always located at odd multiples of half flux quantum for whatever ratio of B/A. There the slope of $E_{J,2\pi}$ is zero, so that the physics is dominated by $E_{J,4\pi}$.

The crossing of the two branches allows the happening of quasiparticle poisoning, which will be right at odd multiples of half flux quantum.

Given the fact that the supercurrent is proportional to the first-order derivative of the energy, quasiparticle-poisoning-induced branch-switching causes no effect on the 2π-period component of supercurrent, but reverses the 4π-period supercurrent.

## 9. Evolution of the phase and the minigap with magnetic flux, more details

In the following we give more plots to illustrate the evolution of the phase (thus the gap) with magnetic flux in the ring and in the junction area, in the presence of the fractional modes.

We note that a flux of $\phi_0/2$ corresponds to a phase of $\pi$ if $\alpha=1$, or to $\pi/2$ if $\alpha=1/2$. In the absence of the $4\pi$-period modes, the $2\pi$-period mode should undergo gap-closing at odd multiples of $\phi_0/2$. In the presence of the $4\pi$-period modes, however, the gap actually does not close at odd multiples of $\phi_0/2$ when the magnetic flux in the junction is less than $\phi_0/2$ [i.e., in the regime of $|B|<3.8$ Gauss in Figs. 2(B) and 2(C)]. Nevertheless, phase jumping still occurs in this regime (see Fig. 2 of the main manuscript), indicating the occurrence of quasiparticle poisoning due to the existence of two $4\pi$-period branches of energy-flux relations and their degeneracy right at odd multiples of $\phi_0/2$, as discussed in the previous section.

As to the positions of the Majoranas, we anticipate that Majoranas might be delocalized inside the junction in the regime of $|B|<3.8$ Gauss [e.g., for the case illustrated in Fig. S9-1(C)] if they exist. In this regime the junction could be regarded as a short junction, as treated by Wieder, Zhang and Kane [3]. Although the minigap in this regime is not closed at odd multiples of half flux quantum because of the half phase-to-flux ratio, the two $4\pi$-period EPRs of the system still cross with each other (as discussed in Section 8 and shown in Fig. S8), yielding superpositions of even and odd-parity branches just like the superposition of occupied and unoccupied quasiparticle states. Quasiparticle poisoning could still happen at the degenerate points, as has been observed.

In the regime of $|B|>3.8$ Gauss, the increased flux in junction tilts the phase-position curves, letting them to reach odd multiples of $\pi$ where the minigap is closed [see Figs. S9-2(C)d, S9-3(C), S9-4(C)], so that Majoranas as superpositions of occupied and unoccupied states become restricted in spaces in the junction, as discussed by Potter and Fu [4]. The positions of the Majoranas are marked by the red dots in Figs. 3 (B) to (E) in the main manuscript, and also in the following figures.

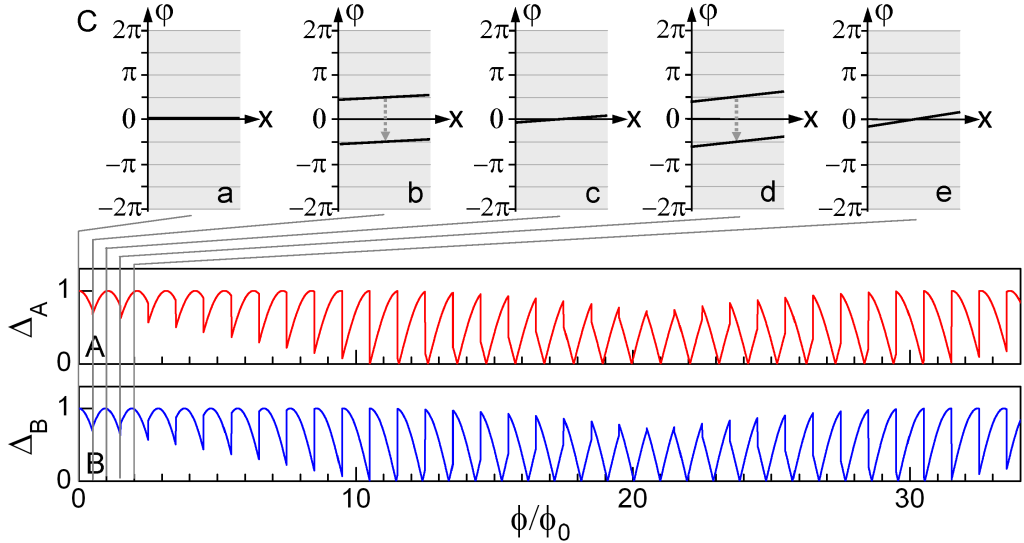

**FIG. S9-1 |** (**A**) and (**B**) Gap energy at positions A and B. (**C**) **a** to **e** Phase distribution in the junction, at ring flux indicated by the gray lines. $\varphi$ may differ from the real value by integer multiples of $2\pi$ depending on the details of quasiparticle poisoning.

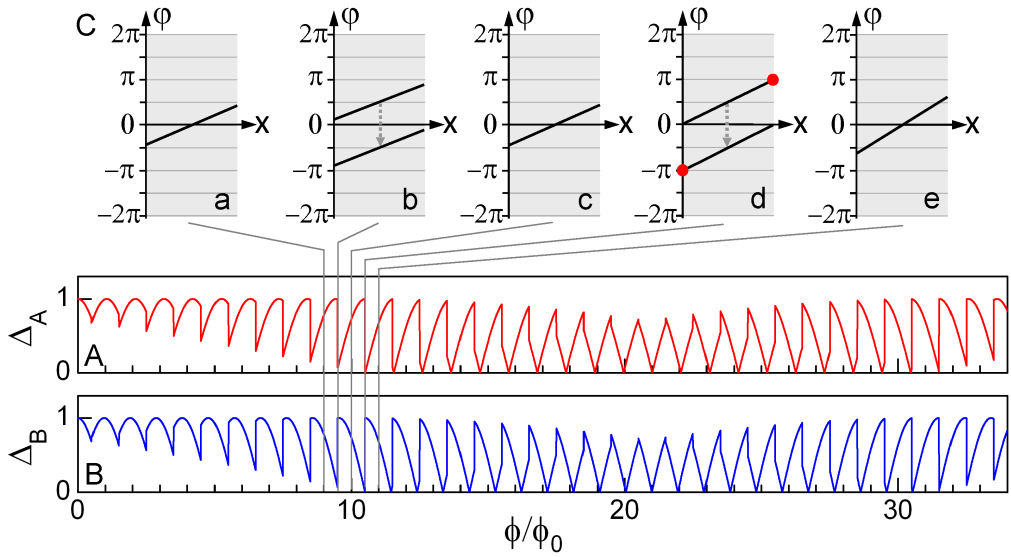

**FIG. S9-2 |** (**A**) and (**B**) Gap energy at positions A and B. (**C**) **a** to **e** Phase distribution in the junction, at ring flux indicated by the gray lines. φ may differ from the real value by integer multiples of 2π depending on the details of quasiparticle poisoning. In **d**, Majoranas start to be restricted in the junction whose positions are marked by the red dots.

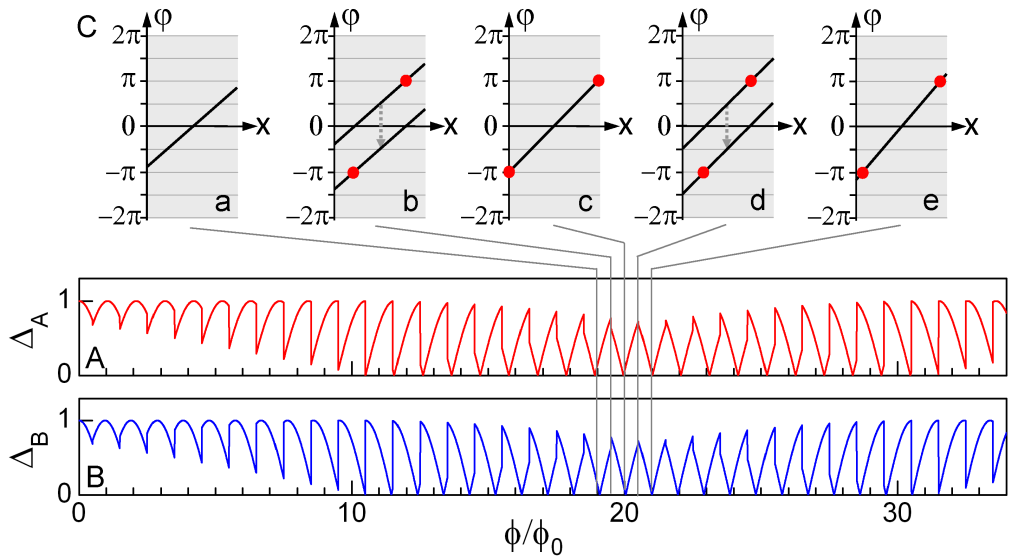

**FIG. S9-3 |** (**A**) and (**B**) Gap energy at positions A and B. (**C**) **a** to **e** Phase distribution in the junction, at ring flux indicated by the gray lines. φ may differ from the real value by integer multiples of 2π depending on the details of quasiparticle poisoning.

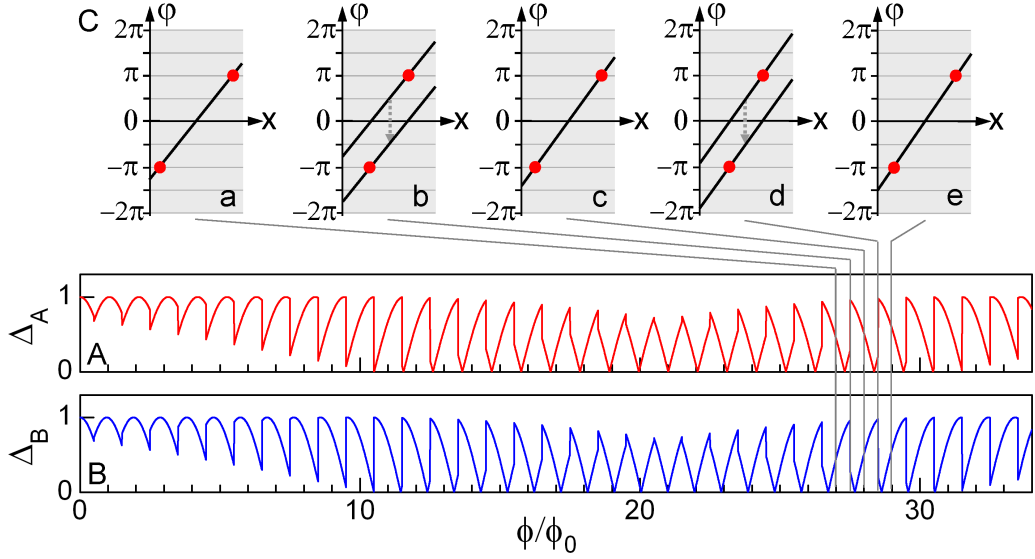

**FIG. S9-4 | (A)** and **(B)** Gap energy at positions A and B. **(C) a** to **e** Phase distribution in the junction, at ring flux indicated by the gray lines. φ may differ from the real value by integer multiples of 2π depending on the details of quasiparticle poisoning.

## 10. Detailed comparison between the data and the model

The minigap shown in Fig. 3 of the main manuscript is obtained with no fitting parameters except for letting α=1/2 and β=1. For a more detailed comparison between the data and the model, we need to put in some fitting parameters.

In the following, let us fit the boundary of the low-resistance state in the 2D plot of $dV/dI_b$ with our model. We assume that the boundary corresponds to a total characteristic current $I_c'$ which contains the contributions of an excess current $I_e$ ($I_e \propto \Delta$), and an empirical leakage current (being a constant, presumably leaking into the bulk superconducting state):

$$I_c' / I_{c0}' = a\Delta/\Delta_0 + b$$

where $I_{c0}'$ is the maximum value of total characteristic current, $\Delta_0$ is the maximum value of minigap, a and b are fitting parameters.

By taking a/b = 3 and using the minigap presented in Fig. 3g of the main manuscript, we obtained the yellow curve in Fig. S10. The agreement between the data and the model seems remarkably well.

In the main manuscript, nevertheless, we prefer just to present the Δ vs. flux curves in Fig. 3 for simplicity.

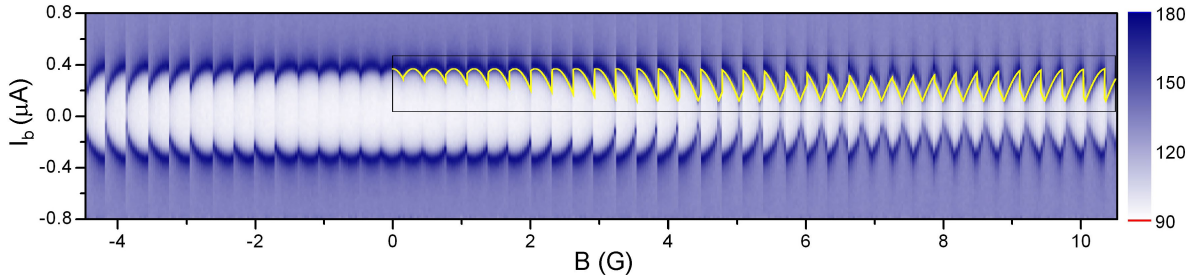

**FIG. S10 |** Comparison between the model (the yellow line) and the boundary of the low-resistance state shown in Fig. 2c of the main manuscript.

11. On the position uncertainty of the Majoranas and the flux uncertainty at which the jumps happen

(1) On the position uncertainty of the Majoranas

In Wieder et al.'s theory [3] the flux in the junction area is not considered (i.e., in the low-field limit, or in the small junction approximation). In this case the Majorana is presumably delocalized inside of the junction area.

In Potter and Fu's theory [4], on the contrary, the flux in the junction area is considered (i.e., in the large-field limit, or in the large junction approximation). The Majoranas are believed to exist at places where the local phase difference crosses $\pi$ such that the minigap is closed. However, due to thermal smearing, these places have an uncertainty. From the data of the device shown in Fig. 2 of the main manuscript we know that its minigap oscillate between 0 and ~28 $\mu V$. Assume a thermal energy of $\Delta E_T = 2$ $\mu V$ at an electron temperature of 20 mK, the phase uncertainty $2\delta$ can be estimated from the relation $\Delta E_T = 28\mu V |\cos[(\pi-\delta)/2]|$. We got $\delta = 0.046\pi$. Such a phase uncertainty corresponds to a length scale uncertainty of the Majorana $\Delta x = 2\delta W = 2 \times 0.046 \times 2.2$ $\mu m = 0.2$ $\mu m$ when the phase difference between A and B is around $\pi$ (e.g., for the case of Fig. 3c of the manuscript). At higher flux the position uncertainty becomes smaller, because of the larger slope of the phase-position curves.

(2) On the flux uncertainty at which the jumps take place

As we have pointed out in Section 8 of the supplementary materials, quasiparticle poisoning will happen at odd multiples of half flux quantum where the even- and odd-parity branches cross with each other. Since the energy scale of the two branches is not but the minigap, the phase uncertainty $2\delta = 2 \times 0.046\pi = 0.092\pi$ also defines a flux uncertainty window of $\sim 0.1\phi_0$ within which quasiparticle poisoning (thus jumping) can take place. From our data, however, the uncertainty window for the happening of quasiparticle poisoning appears to be smaller (being a few tenth of $0.1\phi_0$), which is understandable because in our measurement the flux is ramped slowly along one direction.

12. Why is the $4\pi$-period energy-phase relation the dominant signal measured? The role of the minigap of the surface states versus the superconducting gap of the bulk states

In our $Bi_2Te_3$ samples there are both helical surface states and bulk states.

The helical electrons on $Bi_2Te_3$ surface have relatively long mean-free-path, and presumably with protected transmission at the Pb-$Bi_2Te_3$ interfaces, so that they undergo multiple Andreev reflections between the two Pb electrodes of the Josephson junction. The yielded local Andreev bound states further form quasi 1D structures along the width direction of the junction. The lowest Andreev bound state defines the size of the minigap. And the minigap is a function of position and magnetic flux in the junction.

The bulk states also become superconducting due to the proximity effect at low temperatures, but forming no Andreev bound states at finite energies in the energy window of our experiment, as reflected by the experimental data, presumably because of their trivial nature. They would remain to be superconducting while the minigap of the surface state is tuned to be

closed by the magnetic flux in the junction.

The contact resistance measurement in our experiment is most sensitive to the surface state of the TI. It mainly probes the oscillation and closing of the minigap of the lowest Andreev bound state on the TI surface, despite that the bulk superconducting gap and the related $2\pi$ supercurrent might response differently. This probably explains why a fully skewed EPR is observed in our contact resistance measurement, but a fully skewed CPR was not observed in previous interference/Fraunhofer pattern measurements for critical supercurrent, of which the bulk superconducting state also contributes.

## 13. On the three-terminal measurement configuration for contact resistance measurement

In Fig. S13 we re-plot the rf-SQUID which shows the positions of electrodes A, B, C, and E. We have also sketched on top of the figure the current flowing paths (the red lines) and voltage measurement path (the green lines) in a three-terminal measurement configuration.

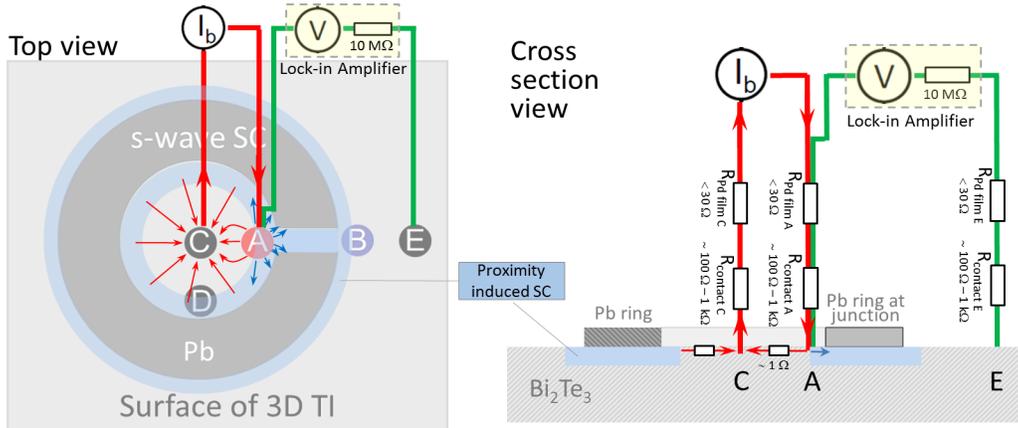

**FIG. S13| Left panel**: Current flowing and voltage drop in our three-terminal measurement configuration. The red lines represent the flowing of normal current, and the blue lines represent the flowing of supercurrent. The current injected from electrode A tends to flow into the superconducting ring, redistributing on the ring, then flowing from the ring to electrode C. The green lines represent the voltage measurement loop. **Right panel**: the equivalent circuit in a cross-section view.

What is the signal we measured in the three-terminal configuration?

Our general picture is: the flux determines the local phase difference in the Pb-Bi$_2$Te$_3$-Pb junction, and the local phase difference determines the local supercurrent density (via the Josephson equation) as well as the local minigap in the junction. The local minigap further determines the measured contact resistance $dV/dI_b$ at the Bi$_2$Te$_3$-Pd interface.

From the right panel of Fig. S13 it can been seen that along the voltage measurement loop there are voltage drops: (1) on $R_{Pd\_film,A}$ which is <30 Ω and being a constant; (2) on $R_{Pd\ contact,A}$ which ranges from ~ 100 Ω to several kΩ in our experiments, and partially oscillating with the flux; and (3) on Bi$_2$Te$_3$ surface which is small, with an equivalent non-local resistance of only ~ 1 Ω (see Y. Pang, et al., arXiv:1503.00838v1). In such case, the dominant oscillatory signal we measured in the three-terminal configuration, with oscillation amplitude of ~ 10 Ω along the line

cut at zero-bias current and up to several tens $\Omega$ along line cuts at high bias currents, can only come from the oscillation of the contact resistance.

Indeed, in the beginning of the experiment we have checked the contact resistance measurement with different combinations of electrodes, such as to measure the contact resistance of electrode A with whatever electrodes C, D, E or even the ring as the second and third electrodes, the semilunar shape and jumping looked to be identical.

## 14. Notes on the measurement currents

In our experiment, the ac excitation current used to measure the $dV/dI_b$ is around 1 nA, being four order of magnitude smaller than the $I_c$ of the Pb-$Bi_2Te_3$-Pb junction (several 10 $\mu A$).

The ramping range of dc bias current $I_b$ was ±0.1 to ±1 $\mu A$, about two order of magnitude smaller than the $I_c$ of the Pb-$Bi_2Te_3$-Pb junction which was a few tens $\mu A$ for the device shown in Fig. 2 of the manuscript. Therefore, the ac and dc currents used in our measurement would not influence the status of the Pb-$Bi_2Te_3$-Pb junctions.

Further increasing $I_b$ to the $I_c$ of Pb-$Bi_2Te_3$-Pb junction will ultimately influence the status of the junction, driving it to the normal state. We believe that this is the reason why we observed a Fraunhofer-like pattern of contact resistance on one of the graphite-based devices, where we used the superconducting ring as one of the current leads.

With the ramping of $I_b$, the applied bias current eventually exceeds the local characteristic supercurrent (several 0.1 $\mu A$) of $Bi_2Te_3$ which is determined by the minigap of the junction. In this way the amplitude of the local minigap is measured.

## 15. Estimation of the effective junction area in the presence of flux compression, stray supercurrent and proximity-effect-induced superconductivity on $Bi_2Te_3$ surface

In Fig. S15 we illustrate how the effective junction area of the device shown in Fig. 2 of the main manuscript is estimated in the presence of flux compression, stray supercurrent distribution and proximity-effect-induced superconductivity on $Bi_2Te_3$ surface. Our previous study shows that the proximity-effect-induced superconductivity spreads from the Pb electrodes to a distance of micron on the surface of $Bi_2Te_3$. (Qu, F. et al., Sci. Rep. 2, 339 (2012); Yang, F. et al., Phys. Rev. B 86, 134504 (2012)). For this reason, at the two ends of the junction there will be areas distributed with stray supercurrents on the surface of $Bi_2Te_3$. Since the strength of superconductivity in these stray areas decays with the distance, their edges are not well defined. Adding to the complexity, the effective area might also vary with the local supercurrent density. Nevertheless, with our experience on the Fraunhofer pattern of single Josephson junctions of the same kind and with similar size, we came up with the empirical method and understanding as shown in the right panel of Fig. S15 for estimating the approximated effective area of the junctions. The error is less than 20%.

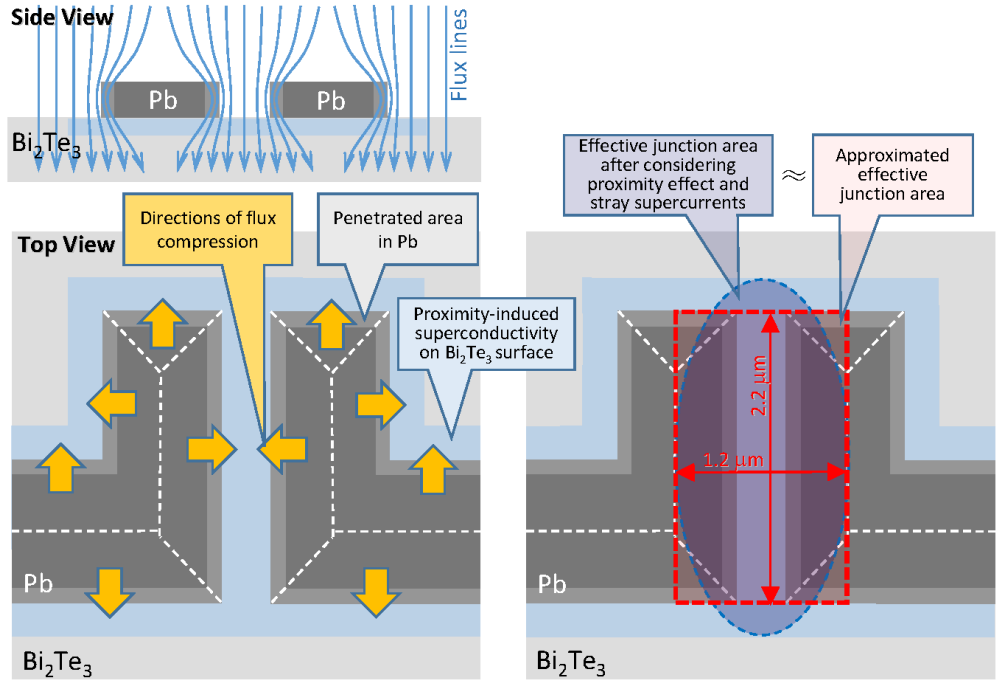

**FIG. S15| Upper left panel**: Illustration of flux compression. **Lower left panel**: The flux within each area defined by the white dashed lines in Pb will be compressed along the arrow direction. **Right panel**: Estimation on the effective junction area in the presence of flux compression and stray supercurrent distribution on $Bi_2Te_3$ surface. The effective junction area for the device shown in Fig. 2 of the main manuscript can be approximately represented by the red box, which is 1.2×2.2 $\mu m^2$.

References:

[1] F. Yang et al., Proximity effect at superconducting Sn-Bi$_2$Se$_3$ interface. *Phys. Rev. B* **85**, 104508 (2012).

[2] A. Barone, "Physics and application of the Josephson effect", John Wiley and Sons, Inc. (1982).

[3] B. J. Wieder, F. Zhang, C. L. Kane, Signatures of Majorana fermions in topological insulator Josephson junction devices, *Phys. Rev. B* **89**, 075106 (2014).

[4] A. C. Potter, L. Fu, Anomalous supercurrent from Majorana states in topological insulator Josephson junctions, *Phys. Rev. B* **88**, 121109 (2013).




\begin{thebibliography}{10}

\bibitem{Majorana} Majorana, E. Teoria simmetrica dellelettrone e del positrone. {\it Nuovo Cimento} {\bf 14}, 171-184 (1937).

\bibitem{Majorana_return_Wilczek} Wilczek, F. Majorana returns. {\it Nature Phys.} {\bf 5}, 614-618 (2009).

\bibitem{Majorana_return_Service} Service, R. F. Search for Majorana fermions nearing success at last? {\it Science} {\bf 332}, 193-195 (2011).

\bibitem{Majorana_return_Franz} Franz, M. Race for Majorana fermions. {\it Physics} {\bf 3}, 24 (2010); Franz, M. Majorana fermions: the race continues. arXiv:1302.3641v1.

\bibitem{Fu_liang_2008} Fu, L., Kane, C. L. Superconducting proximity effect and Majorana fermions at the surface of a topological insulator. {\it Phys. Rev. Lett.} {\bf 100}, 096407 (2008).

\bibitem{Kowenhoven} Mourik, V. {\it et al.} Signatures of Majorana fermions in hybrid superconductor-semiconductor nanowire devices. {\it Science} {\bf 336}, 1003-1007 (2012).

\bibitem{Kitaev2001} Kitaev, A. Yu. Unpaired Majorana fermions in quantum wires. {\it Phys.-Usp.} {\bf 44}, 131-136 (2001).

\bibitem{Fu_Kane_2009ring} Fu, L., Kane, C. L. Josephson current and noise at a superconductor/quantum-spin-Hall-insulator/superconductor junction. {\it Phys. Rev. B} {\bf 79}, 161408 (2009).

\bibitem{Lutchyn_Sau_Das_Sarma_2010ring} Lutchyn, R. M., Sau, J. D., Das Sarma, S. Majorana fermions and a topological phase transition in semiconductor-superconductor heterostructures. {\it Phys. Rev. Lett.} {\bf 105}, 077001 (2010).

\bibitem{Beenakker_2013ring} Diez, M. {\it et al.} Phase-locked magnetoconductance oscillations as a probe of Majorana edge states. {\it Phys. Rev. B} {\bf 87}, 125406 (2013).

\bibitem{Benjamin_Zhang_Kane_2014ring} Wieder, B. J., Zhang, F., Kane, C. L. Signatures of Majorana fermions in topological insulator Josephson junction devices. {\it Phys. Rev. B} {\bf 89}, 075106 (2014).

\bibitem{Brinkman} Veldhorst, M. {\it et al.} Josephson supercurrent through a topological insulator surface state. {\it Nature Mater.} {\bf 11}, 417-421 (2012); Veldhorst, M. {\it et al.} Experimental realization of superconducting quantum interference devices with topological insulator junctions. {\it Appl. Phys. Lett.} {\bf 100}, 072602 (2012).

\bibitem{QuFM} Qu, F. {\it et al.} Strong superconducting proximity effect in Pb-Bi$_2$Te$_3$ hybrid structures. {\it Sci. Rep.} {\bf 2}, 339 (2012).

\bibitem{Goldhaber-Gordon} Williams, J. R. {\it et al.} Unconventional Josephson effect in hybrid superconductor-topological insulator devices. {\it Phys. Rev. Lett.} {\bf 109}, 056803 (2012).

\bibitem{Rokhinson} Rokhinson, L. P. {\it et al.} The fractional a.c. Josephson effect in a semiconductor-superconductor nanowire as a signature of Majorana particles. {\it Nature Phys.} {\bf 8}, 795-799 (2012).

\bibitem{HarlingenTI} Kurter, C. {\it et al.} Dynamical gate-tunable supercurrents in topological Josephson junctions. {\it Phys. Rev. B} {\bf 90}, 014501 (2014); Kurter, C. {\it et al.} Evidence for an anomalous current-phase relation in topological insulator Josephson junctions. {\it Nature Commun.} {\bf 6}, 7130 (2015).

\bibitem{Moler_PRL} Sochnikov, I. {\it et al.} Nonsinusoidal current-phase relationship in Josephson junctions from the 3D topological insulator HgTe. {\it Phys. Rev. Lett.} {\bf 114}, 066801 (2015).

\bibitem{Tarucha_Molenkamp} Wiedenmann, J. {\it et al.} Zero-energy Andreev bound states in a HgTe-based topological Josephson junction. arXiv:1503.05591.

\bibitem{Yacoby_NW} Hart, S. {\it et al.} Induced superconductivity in the quantum spin Hall edge. {\it Nature Physics} {\bf 10}, 638-643 (2014).

\bibitem{Yacoby_QPP} Lee, S. P. {\it et al.} Revealing topological superconductivity in extended quantum spin Hall Josephson junctions. {\it Phys. Rev. Lett.} {\bf 113}, 197001 (2014).

\bibitem{Kouwenhoven_Qu} Pribiag, V. S. {\it et al.} Edge-mode superconductivity in a two dimensional topological insulator. {\it Nature Nanotechnology} {\bf 10}, 593-597 (2015).

\bibitem{Yang_Fan2} Yang, F. {\it et al.} Proximity-effect-induced superconducting phase in the topological insulator Bi$_2$Se$_3$. {\it Phys. Rev. B} {\bf 86}, 134504 (2012).

\bibitem{BTK} Blonder, G. E., Tinkham, M., Klapwijk, M. T. Transition from metallic to tunneling regimes in superconducting microconstrictions: excess current, charge imbalance, and supercurrent conversion. {\it Phys. Rev. B} {\bf 25}, 4515 (1982).

\bibitem{Josephson} Barone, A. {\it Physics and application of the Josephson effect}. John Wiley and Sons, Inc. (1982); Golubov, A. A., Kupriyanov, M. Yu., Il¡¯ichev, E. The current-phase relation in Josephson junctions. {\it Rev. Mod. Phys,} {\bf 76}, 411 (2004) and the references therein.

\bibitem{HJLee} Lee, G.-H. {\it et al.} Ultimately short ballistic vertical graphene Josephson junctions. {\it Nature Commun.} {\bf 6}, 6181 (2015).

\bibitem{Yang_Fan} Yang, F. {\it et al.} Proximity effect at superconducting Sn-Bi$_2$Se$_3$ interface. {\it Phys. Rev. B} {\bf 85}, 104508 (2012).

\bibitem{QP_poisoning_Loss} Rainis, D., Loss, D. Majorana qubit decoherence by quasiparticle poisoning. {\it Phys. Rev. B} {\bf 85}, 174533 (2012).

\bibitem{QP_poisoning_Beenakker_2013} Beenakker, C. W. J. {\it et al.} Fermion-parity anomaly of the critical supercurrent in the quantum spin-Hall effect. {\it Phys. Rev. Lett.} {\bf 110}, 017003 (2013).

\bibitem{Fu_SJJ_2013} Potter, A. C., Fu, L. Anomalous supercurrent from Majorana states in topological insulator Josephson junctions. {\it Phys. Rev. B} {\bf 88}, 121109 (2013).

\bibitem{Wilczek} Arovas, D., Schrieffer, J. R. and Wilczek, F. {\it Phys. Rev. Lett.} {\bf 53}, 722 (1984).

\end{thebibliography}
\end{document}